\documentclass[aps,prl,twocolumn,superscriptaddress,nobalancelastpage,amsmath,amssymb,
longbibliography]{revtex4-2}
\usepackage[utf8]{inputenc}
\usepackage{float}
\usepackage{amsmath}
\usepackage{tikz}
\usepackage{graphicx}
\usepackage{dcolumn}
\usepackage{multirow}
\usepackage{comment}

\usepackage{braket}

\def\[#1\]{%
  \begin{equation}#1\end{equation}%
}

\usepackage{framed} 
\definecolor{shadecolor}{gray}{0.95}

\usepackage{bm}

\usepackage{hyperref}
\hypersetup{
    unicode=false,          
    pdftoolbar=true,        
    pdfmenubar=true,        
    pdffitwindow=false,     
    pdfstartview={FitH},    
    pdfsubject={},   
    pdfcreator={},   
    pdfproducer={}, 
    pdfkeywords={} {} {}, 
    pdfnewwindow=true,      
    colorlinks=true,       
    linkcolor=blue, 
    citecolor=blue,        
    filecolor=blue,      
    urlcolor=blue           
}

\newcommand{\tpar}{t_{\parallel}}
\newcommand{\tper}{t_{\perp}}

\begin{document}

\title{Gapless and Non-Equilibrium Topology in Blockaded Ladders}

\author{Ramanjit Sohal}
\affiliation{Pritzker School of Molecular Engineering, University of Chicago, Chicago, IL 60637, USA}

\author{Hannes Bernien}
\affiliation{Institute for Quantum Optics and Quantum Information of the Austrian Academy of Sciences, 6020 Innsbruck, Austria}
\affiliation{Institute for Experimental Physics, University of Innsbruck, 6020 Innsbruck, Austria}

\author{Ruben Verresen}
\affiliation{Pritzker School of Molecular Engineering, University of Chicago, Chicago, IL 60637, USA}

\begin{abstract}
We propose a multi-leg PXYP model as a minimal setting to explore constraint-induced intertwinement between itinerant and internal degrees of freedom.
We argue that the Rydberg-like blockade induces correlations in \emph{squeezed space}, providing a unified mechanism for nontrivial string order both in and out of equilibrium.
In a two-leg geometry, this leads to an \emph{intrinsically} gapless topological ground state, which can terminate at a supersymmetric multicritical point.
This mechanism also leads to the preservation of string order for parametrically long times after a quench, which we verify using tensor network simulations and a classical stochastic approximation to the dynamics.
In a three-leg geometry, the blockade leads to an additional emergent $U(1)$ symmetry, resulting in a multicomponent Luttinger liquid where both itinerant and internal degrees of freedom are critical. This family of models arises naturally in a range of  quantum simulator platforms such as Rydberg tweezer arrays and dipolar atoms in optical lattices. 
\end{abstract}

\date{\today}
\maketitle

\textit{Introduction---}%
\emph{Constraints} in many-body systems have long been known to be a route to a host of exotic phenomena, both in and out of equilibrium. Indeed, constrained systems often exhibit \emph{topological} physics, with the paradigmatic example being provided by quantum dimer models, originally studied in the context of high-$T_c$ superconductors, but subsequently understood to be theoretically tractable arenas in which to study spin liquid physics \cite{Rokhsar1988,Moessner2001,Misguich2002,Fradkin2013}. Conversely, it is natural to expect that constraints may inhibit the ergodicity of a many-body system driven out of equilibrium. 
Indeed, it is now well-established that multiple routes to evading thermalization, such as quantum many-body scars and Hilbert space fragmentation, naturally arise in such constrained systems \cite{Turner2018a,Turner2018b,Sala2020,Khemani2020,Serbyn2021,Moudgalya2022a}.

Recent interest in constrained systems has  been driven by the development of quantum simulators in which such systems are natively realized \cite{Browaeys2020}.
For instance, the observation of peculiar non-thermal behavior in Rydberg atom tweezer arrays was later understood to be 
a dynamical consequence of quantum many-body scars in the PXP model \cite{Fendley2004,Lesanovsky2012,Bernien2017,Turner2018a,Turner2018b}. The Rydberg blockade has also provided a natural ingredient with which to realize the aforementioned quantum dimer models, underlying the proposals for emergent gauge theories \cite{Glaetzle2014, Celi2020, Surace2020, Verresen2021a, Samajdar2021, Giudici2022, Tarabunga2022, Verresen2022, Samajdar2023, Sahay2023, Shah2025, Zeng2025} leading to experimental signatures of $\mathbb{Z}_2$ \cite{Semeghini2021} and $U(1)$ spin liquids \cite{Geim2026} (see also \cite{Karch2026,Bornet2026}).

\begin{figure}
    \centering
    \includegraphics[width=0.99\linewidth]{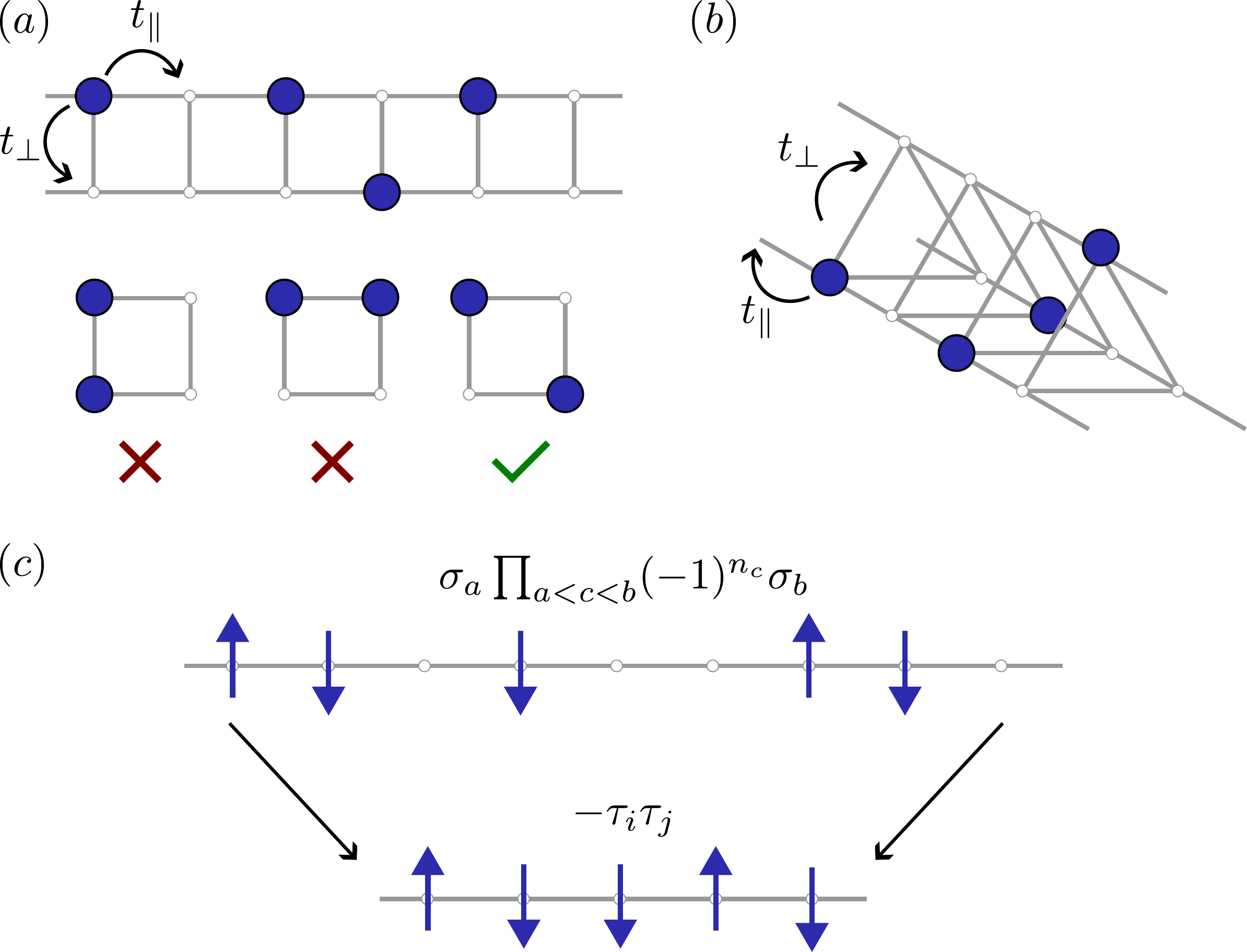}
    \caption{(a) Two-leg PXYP model and blockaded configurations; blue circles represent hardcore bosons. (b) Three-leg PXYP model. (c) Construction of squeezed space states in the two-leg model, with the leg index represented as a pseudospin. String order in the physical space maps to conventional correlators in squeezed space.}
    \label{fig:model}
\end{figure}

Given the experimental relevance of such constrained systems, this raises the question: can one find a class of constrained models for which the constraint provides a \textit{common} origin for both nontrivial equilibrium and nonequilibrium properties? 
We answer this question in the affirmative, using as representatives \emph{multi-leg PXYP ladder models}---XY models defined on a multi-leg ladder, subject to the Rydberg blockade, as depicted in Fig.~\ref{fig:model}, which generalize single chain PXYP models studied previously \cite{Singh21,Yang25,Liu2025,Corcoran25,Naz24}. 
We argue that the Rydberg blockade induces correlations between \emph{internal} and \emph{itinerant} degrees of freedom (i.e. degrees of freedom perpendicular and parallel, respectively, to the length of the ladder), leading to \emph{string order} both in and out of equilibrium. In a two-leg geometry, we show this leads to gapless, topological ground states \cite{Kestner2011,Fidkowski2011,Cheng2011,Sau2011,Keselman2015,Scaffidi2017,Verresen2018,Parker2018,Verresen2021,Thorngren2021,Li2024SPT,Wen2025}
and the parametrically slow decay of string order after a quench. In a three-leg geometry, this leads to an emergent $U(1)$ symmetry and a multi-component Luttinger liquid with quasi-long range string order. 

These PXYP models are closely related to the effective description of Rydberg tweezer arrays as a PXP model in the large detuning limit as well as dipolar cold atom systems in the limit of large interactions. 
We compare the physics of these systems with that of the PXYP models, finding qualitatively similar behavior. 
Remarkably, along the way, we find that the PXP model in the large detuning limit can naturally exhibit \emph{supersymmetry}.  
The PXYP models are thus minimal in a dual sense: they provide a theoretically minimal example of the constraint-induced intercorrelation between itinerant and internal degrees of freedom discussed above, and are closely related to experimental settings.

\textit{Multi-Leg PXYP Model---}%
The PXYP model is a generalization of the celebrated PXP model, defined on an arbitrary lattice of qubits,
\begin{align}
H_{\mathrm{PXYP}} = &- t \sum_{\langle i,j\rangle} P \left( \sigma^+_i \sigma^-_j + h.c. \right) P  \, ,\label{eq:PXYP}
\end{align}
where $P$ implements the nearest-neighbor ``Rydberg blockade": $n_i n_j = 0$. While there has been growing interest in this natural and rich model \cite{Singh21,Yang25,Liu2025,Corcoran25,Naz24} (with the one-dimensional case being an integrable critical chain \cite{Alcaraz99}), here we will see how new physics emerges by generalizing to the case of an \textit{$N$-leg ladder}; as examples, we will focus on $N=2$ and $N=3$ 
[see Figs.~\ref{fig:model}(a,b)]. 
In such a geometry, the model has a $G = U(1) \times \mathbb Z_N$ symmetry corresponding to the $U(1)$ generator $Q = \frac{1}{2} \sum_i \sigma^z_i$ and the 
$\mathbb{Z}_N$ rotation of the ladder which cyclically permutes the legs of the ladder.
If we interpret Eq.~\eqref{eq:PXYP} as a model of hardcore bosons, as is customary in Rydberg-blockaded models, and treat the $N$ sites connected by rungs as a single unit cell, then the $U(1)$ symmetry corresponds to particle-number conservation and the $\mathbb{Z}_N$ rotation may be viewed as an internal symmetry. The Rydberg blockade thus naturally intertwines the former, \emph{itinerant} degrees of freedom with the latter, \emph{internal} degrees of freedom. It is in this sense that Eq.~\eqref{eq:PXYP} provides a minimal realization of the constrained physics advertised above.

\textit{Experimental Realizations---}%
The PXYP model of Eq.~\eqref{eq:PXYP} is closely related to a particular limit of the celebrated PXP model \cite{Fendley2004,Lesanovsky2012,Turner2018a}: 
\begin{equation}
H_{\mathrm{PXP}} = -\frac{\Omega}{2}\sum_j  P \sigma^x_j P - \delta \sum_j n_j. \label{eq:PXP}
\end{equation}
For large detuning $\delta$, sectors with different particle number $\sum_j n_j$ become energetically decoupled. At second-order in perturbation theory in $\Omega/\delta$, $H_{\mathrm{PXP}}$ is approximated by \cite{Yang25},
\begin{align}
    H_{\mathrm{PXYP}}' \equiv  H_{\mathrm{PXYP}} + t\sum_i P_i -(\delta+2t)\sum_i n_i \, , \label{eq:PXP_Eff}
\end{align}
with $t=\frac{\Omega^2}{4\delta}$ and $P_i = \prod_{ \langle j, i \rangle } (1-n_j)$. 
Despite the presence of density-density interactions, we find the physics of Eqs.~\eqref{eq:PXYP} and \eqref{eq:PXP_Eff}  to be qualitatively similar;
in the following, we will focus our analysis on the former due to its simplicity and point out differences with the latter. 
We note this is distinct from Ref.~\cite{Wang2025}, which studied the ground state and dynamics of Eq.~\eqref{eq:PXP} in similar geometries for finite detuning.

We remind the reader that the PXP model, Eq.~\eqref{eq:PXP}, is only an effective description of Rydberg atoms, since there are longer-range interaction tails $\sim 1/r^6$. On a square lattice, if we use a single species of atoms and we call the nearest-neighbor (NN) blockade strength $V$, then the next-nearest-neighbor correction has strength $\frac{V}{\sqrt{2}^6} = \frac{V}{8}$, which can be sizable. In principle this need not affect the long-distance properties, since density-density interactions preserve the $U(1) \times \mathbb Z_N$ symmetry. However, Rydberg interactions can suffer from disorder, which is thus desirable to avoid. This can be achieved with a dual-species set-up where intraspecies interactions are made negligible. 
Placing the atoms in a checkerboard pattern means the next-NN correction is negligible while the fourth-NN interaction gives the leading correction of strength $\frac{V}{\sqrt{5}^6} = 0.008 V$. 

The multi-leg PXYP model also finds natural realizations in, e.g., itinerant dipolar atom systems \cite{Baier2016,Su2023,Su2026}. Such systems can be represented in terms of hardcore bosons and modeled by a Hamiltonian of the form,
\begin{align}
	H_{\mathrm{dipolar}} = -t \sum_{\langle i , j \rangle} (b_i^\dagger b_j + h.c.) + g \sum_{\langle i , j \rangle} n_i n_j,
\end{align}
where $n_i = b_i^\dagger b_i$. In the limit $g \to \infty$, one lands on $H_{\mathrm{PXYP}}$
at lowest order in perturbation theory, expressed in terms of hardcore bosons. Importantly, Eq.~\eqref{eq:PXYP} arises naturally in a limit obtained by tuning a single parameter in systems involving at most two-body interactions.

\textit{Squeezed Space---}%
The separation into itinerant and internal degrees of freedom is made most manifest by the \emph{squeezed space} construction \cite{ogata1990,zhang1997,kruis2004}. 
Given an occupation number basis state, the corresponding squeezed space state is constructed by deleting all the holes, yielding a string of flavor labels of the particles. Let us focus on $N=2$ for concreteness [see Fig.~\ref{fig:model}(c)]; the general case follows readily.
Explicitly, let us label the positions of the particles as $x_1 < x_2 < \dots < x_{N_p}$ and denote their corresponding flavors with $\sigma_j = \pm 1$. 

Flavor correlations in squeezed space manifest as \emph{string order} in the physical space \cite{kruis2004}. Indeed, consider the string order parameter, $S_{ab} = \sigma_a \prod_{a<c<b} (-1)^{n_c} \sigma_b$, which is the conventional two-point correlator dressed by a particle parity string. Let us define $\tau_i = (-1)^i \sigma_i$, which removes the antiferromagnetic staggering in squeezed space. If sites $a$ and $b$ are occupied by particles $i$ and $j$,  
respectively, then 
$S_{ab} = -\sigma_i (-1)^{j-i} \sigma_j = -\tau_i \tau_j$. That is, the string order in physical space simply measures conventional antiferromagnetic (AFM) order in squeezed space. In contrast, the two-point flavor correlator evaluates to $C_{ab}=\sigma_a \sigma_b = S_{ab}(-1)^{N_{ab}}$, where $N_{ab}$ is the number of particles 
between $a$ and $b$. 
Physically, while the motion of itinerant holes can wash out conventional correlations, the parity string in the string order parameter exactly cancels out the hole contributions to isolate the flavor correlations of the particles.
As we will show next, the Rydberg blockade naturally leads to correlations in squeezed space. 

\begin{figure}
    \centering
    \includegraphics[width=\linewidth]{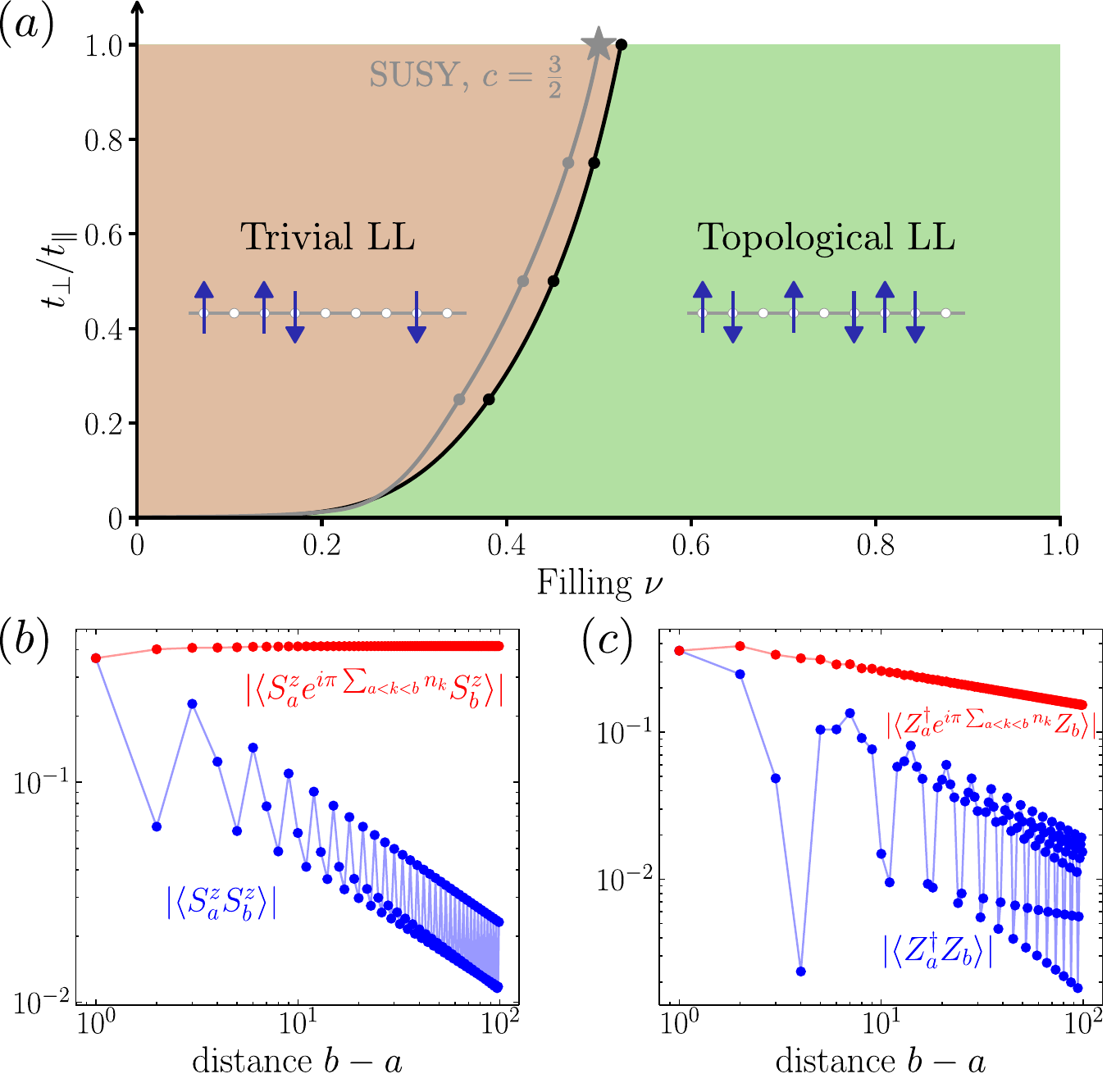}
    \caption{ (a) Phase diagram of the two-leg PXYP model, $H(\tpar,\tper)$.  The solid black line indicates the transition line; the black dots indicate positions of the transition obtained via iDMRG, while the shape at low filling is determined in the End Matter. The grey line/points indicate the transition for $H'(\tpar,\tper)$; the star indicates the SUSY point. The green phase is an intrinsically gapless topological phase. (b) Two-point correlator and string order parameter evaluated using iDMRG  in the igSPT phase of the PXYP model, with $t_\perp = t_\parallel$ and $\nu=2/3$. (c) Two-point correlator and string order parameters for the three-leg model with $t_\perp = 0.2t_\parallel$ and $\nu=6/7$ forming a two-component LL ($c=2$).} 
    \label{fig:igSPT}
\end{figure}

\textit{Two-Leg Gapless Topology---}%
Indeed, focusing first on $N=2$, we show the blockade leads to an intrinsically gapless symmetry-protected topological phase (igSPT) \cite{Thorngren2021} (a gapless state with topologically protected boundary modes and string order) protected by $U(1) \times \mathbb{Z}_2$ symmetry---a symmetry which cannot protect any gapped SPT phase \cite{Pollmann2012,Schuch2011,Chen2011}.
We may represent the leg index as a pseudospin variable, such that the two-leg system may be interpreted as a 1D chain of spinful hardcore particles $a_{j,s}$, where $s \in \{ \uparrow,\downarrow\}$. We study the Hamiltonian,
\begin{equation}
H(t_\parallel,t_\perp)  = - t_\parallel \sum_{j,s} \left( P a_{j,s}^\dagger a_{j+1,s}^{\vphantom \dagger} P +h.c. \right) - t_\perp \sum_j P S^x_j P \label{eq:hamiltonian-base} 
\end{equation}
where $S^\mu_j = a^\dagger_{j,\alpha}\sigma^\mu_{\alpha\beta} a_{j,\beta}$ and $\sigma^\mu$ are the Pauli matrices. 
Note that we have allowed for different couplings for hoppings along the legs and rungs; for $t_\parallel = t_\perp = t$ this reduces to Eq.~\eqref{eq:PXYP}.
The Rydberg constraint may be expressed in this language as $S_j^z S_{j+1}^z = -n_j n_{j+1}$, where $n_j = n_{j,\uparrow} + n_{j,\downarrow}$ is the total particle number on site $j$ \footnote{We note that the statistics does not affect the physics at hand, so $a_{j,s}$ can be taken to be a hardcore boson or fermion.}.
We also consider a similar generalization of Eq.~\eqref{eq:PXP_Eff}, $H'(t_\parallel,t_\perp)$, the explicit form of which we give in the End Matter.

In Fig.~\ref{fig:igSPT}(a) we present the phase diagram of $H(t_\parallel,t_\perp)$ as a function of $t_\perp/t_\parallel$ and filling $\nu$; here $\nu$ is the number of particles per rung. We also compare it with that of $H'(t_\parallel,t_\perp)$.
We argue below that both Hamiltonians support a trivial Luttinger liquid (LL) phase at low filling and a topological LL (an igSPT) at high filling, separated by a phase transition which, at least at low filling, is a $c=3/2$ conformal field theory (CFT).
The physical mechanism underlying the igSPT order is similar to that of the Ising-Hubbard model discussed in Ref.~\onlinecite{Thorngren2021}. At maximal filling the ground state is fixed by the blockade to exhibit AFM order. On doping below maximal filling, the particles become dynamical and form a Luttinger liquid; however, the blockade still encourages AFM order in squeezed space, yielding nontrivial string order.

We gain an analytical handle on this picture by noting that $H(t_\parallel, t_\perp)$ on the line $t_\perp=0$ is exactly solvable. 
The spin degrees of freedom are frozen and the ground state lies in the subspace where the squeezed space configurations exhibit perfect AFM order; the bosons thus behave as nearest-neighbor hopping hardcore particles, realizing a Luttinger liquid \cite{supp}. The perfect squeezed space AFM implies long-range string order, yielding an igSPT for $\tper = 0$.

On increasing $t_\perp$, the AFM order in squeezed space is eventually lost at a critical value of $t_\perp$. 
This picture can be made quantitative in the limit of small $\nu$, as shown in the End Matter, allowing us to determine the shape of the transition line. In this limit, the transition is a CFT with central charge $c=\frac{3}{2}$, corresponding to an Ising CFT, from the critical squeezed space spins, on top of the Luttinger liquid. 
Conversely, for small $\tper$ and close to maximal filling, deviations from AFM cost $O(1)$ energy \cite{supp}, and thus the internal degrees of freedom are effectively \emph{frozen}. The itinerant degrees of freedom thus exhibit the same behavior as at $t_\perp=0$, yielding a Luttinger liquid which, combined with the squeezed space AFM order, produces an igSPT as claimed above. 

We support this picture using iDMRG simulations \cite{White1992,White1993,Hauschild2018,tenpy2024}, which we use to establish the (black) phase transition line in Fig.~\ref{fig:igSPT}(a). 
For $\nu$ above the transition, as shown in Fig.~\ref{fig:igSPT}(b), we see the conventional two-point correlator exhibits quasi-long range order while the string order parameter exhibits genuine long-range order, indicative of an igSPT. 
We confirm the gapless nature of these phases in the End Matter where we extract the central charge using finite-entanglement scaling \cite{Pollmann2009}.

Turning briefly to $H'(t_\parallel,t_\perp)$, 
remarkably, we show in the End Matter that this model is \emph{equivalent} to the supersymmetric (SUSY) fermion chain of Ref.~\onlinecite{Bauer2013} along the $t_\perp=t_\parallel=1$ line. Recall that this corresponds to the effective Hamiltonian of the large-detuning PXP model, Eq.~\eqref{eq:PXP_Eff}, implying lattice SUSY can naturally arise in Rydberg atom arrays. 
At $\nu=\frac{1}{2}$ this model realizes the $\mathcal{N}=2$ SUSY CFT with $c=\frac{3}{2}$, indicated by the star in Fig.~\ref{fig:igSPT}(a). 
In contrast, previous Rydberg-based SUSY proposals study emergent SUSY at a critical point \cite{Li2024,Wang2025,Wang2026} or use Floquet engineering to realize lattice SUSY \cite{Liu2025}.
Along the line $\tper = 0$, we expect the sector with exact AFM squeezed-space order to remain lowest in energy; the density interactions in Eq.~\eqref{eq:PXP_Eff} project to nearest-neighbor density interactions, which are marginal in the Luttinger liquid phase. We thus expect the $\tper = 0$ line to remain an igSPT.
Our numerics suggest the phase diagrams of $H'(t_\parallel,t_\perp)$ and $H(t_\parallel,t_\perp)$ are otherwise qualitatively alike; the grey line in Fig.~\ref{fig:igSPT}(a) indicates the phase transition for the former.

To understand the symmetry protection of the igSPT, we note that the $\mathbb{Z}_2^P$ subgroup of $U(1)$ particle number conservation corresponding to particle number parity is gapped out at low energies; adding or removing a single particle necessarily violates the perfect AFM order in squeezed space. Thus, while the Hamiltonian has the total symmetry $U(1) \times \mathbb{Z}_2$, only the quotient group $(U(1)/\mathbb{Z}_2^P) \times \mathbb{Z}_2$ acts faithfully at low energies. The nontrivial string order indicates this effective symmetry exhibits an \emph{emergent anomaly}, yielding an igSPT \cite{Thorngren2021}.

\textit{Three-Leg String Order---}%
Crucially, this physics is not restricted to the two-leg ladder. Indeed, consider the case of $N=3$, described by Eq.~\eqref{eq:hamiltonian-base} where now $s=0,1,2$ denotes the three internal states, and we replace $S^{x,z} \mapsto X, Z$, the generalized Pauli operators, such that: $X\ket{a} = \ket{a+1 \, \text{mod } 3}$. 
The full symmetry is now $U(1)\times S_3$, where $S_3$ corresponds to the permutations of the three legs.
At maximal filling, there are no itinerant degrees of freedom, while the blockade projects the internal degrees of freedom into the ground state space of a three-state AFM Potts model with no transverse field when $t_\perp=0$, yielding an extensive degeneracy. Turning on $t_\perp$ lifts this degeneracy; projecting Eq.~\eqref{eq:hamiltonian-base} into this subspace yields an XY Hamiltonian in nearest-neighbor flavor differences \cite{supp}, and hence a Luttinger liquid. Remarkably, we thus obtain an emergent $U(1)$ symmetry even though at maximal filling the only microscopic symmetry is the discrete $S_3$.

Now, as in the two-leg case, at small hole doping and $t_\perp$, the itinerant and internal degrees of freedom remain decoupled \cite{supp}. In this regime, the system should therefore realize a two-component Luttinger liquid, arising from the simultaneously gapless itinerant and internal degrees of freedom. We confirm this numerically in the End Matter. A bosonization analysis can be used to verify the perturbative stability of the $c=2$ phase to small doping \cite{supp}. Notably, \emph{both} the conventional correlator and the string order exhibit algebraic decay, with the latter decaying more slowly [see Fig.~\ref{fig:igSPT}(c)]. 
Let us emphasize that the two gapless components do not correspond simply to two decoupled local degrees of freedom; one component describes gapless fluctuations in squeezed space, which are nonlocal in the physical space. This highlights the intertwinement of the internal and itinerant degrees of freedom due to the blockade.

\textit{Nonequilibrium Topology---}%
In general, SPT order---gapless or otherwise---is expected to be a property of the low energy physics of a system. 
It is then natural to expect that under a quench of the PXYP Hamiltonian from a high-energy state, any string order should rapidly decay as the system thermalizes.
Remarkably, focusing on the two-leg case, we find that the Rydberg blockade inhibits the decay of string order after such a quench; though it does inevitably decay to zero, it remains non-zero for parametrically long time scales controlled by the filling fraction.

\begin{figure}
    \centering
    \includegraphics[width=\linewidth]{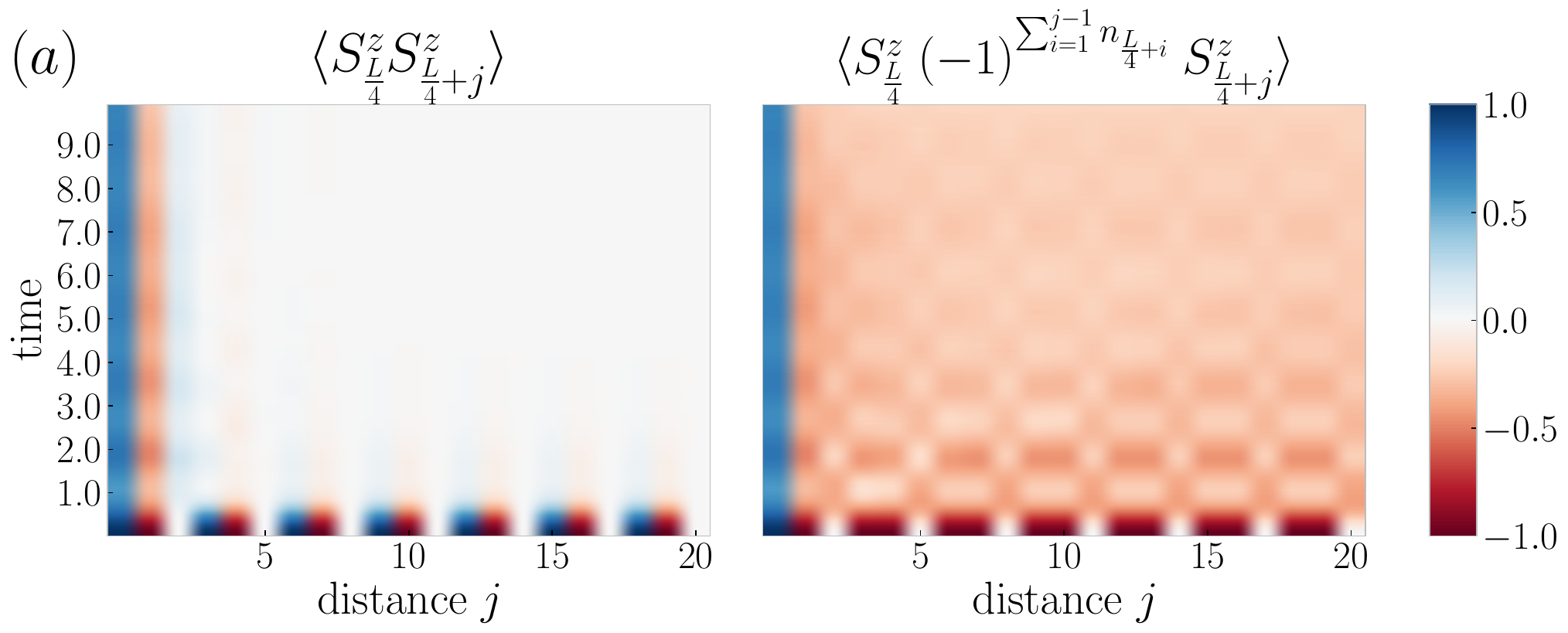}
    \includegraphics[width=\linewidth]{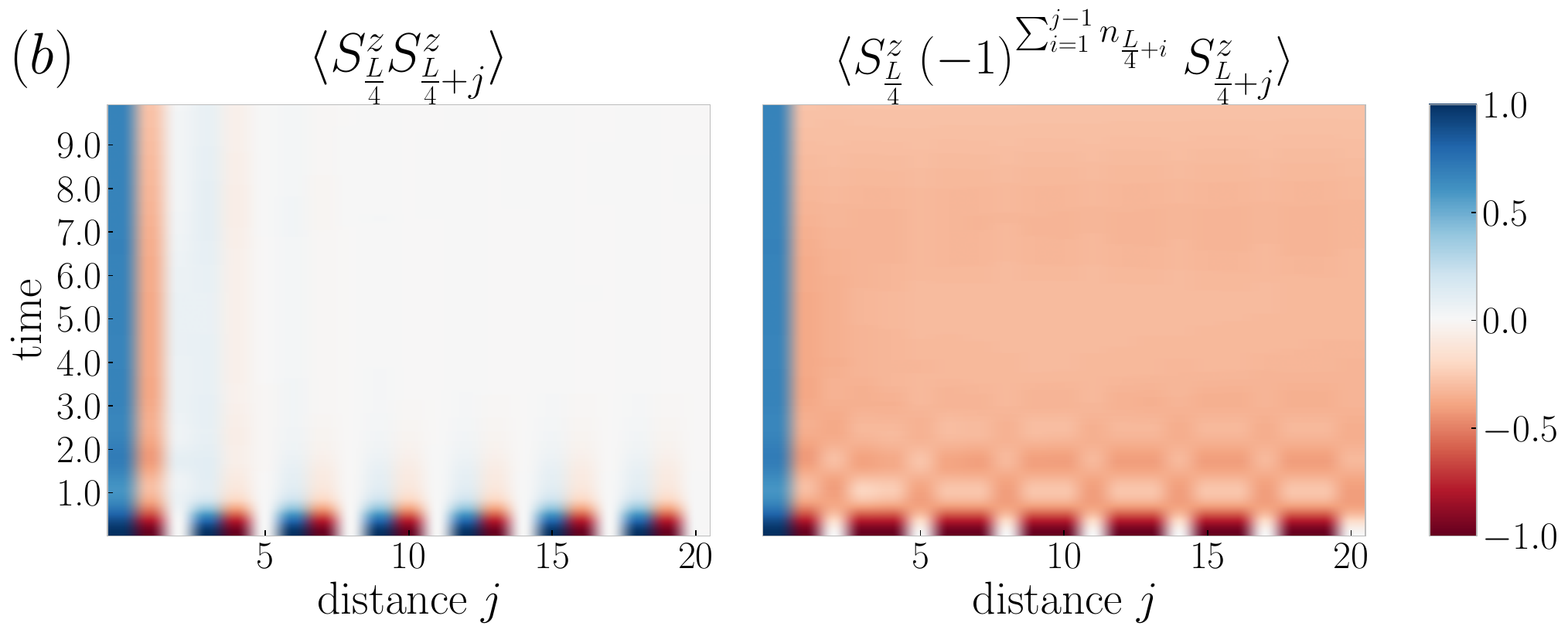}
    \caption{Spatiotemporal dependence of the (left) conventional correlator and (right) string order parameter after a quench from the state $\ket{\uparrow \downarrow 0}^{\otimes L/3}$ with time dependence generated by (a) the PXYP Hamiltonian and (b) the effective large-detuning PXP Hamiltonian. In both cases, we find nonlocal string order persisting significantly longer than local correlations.}
    \label{fig:quantum-quench}
\end{figure}

\begin{figure}
    \centering
    \includegraphics[width=\linewidth]{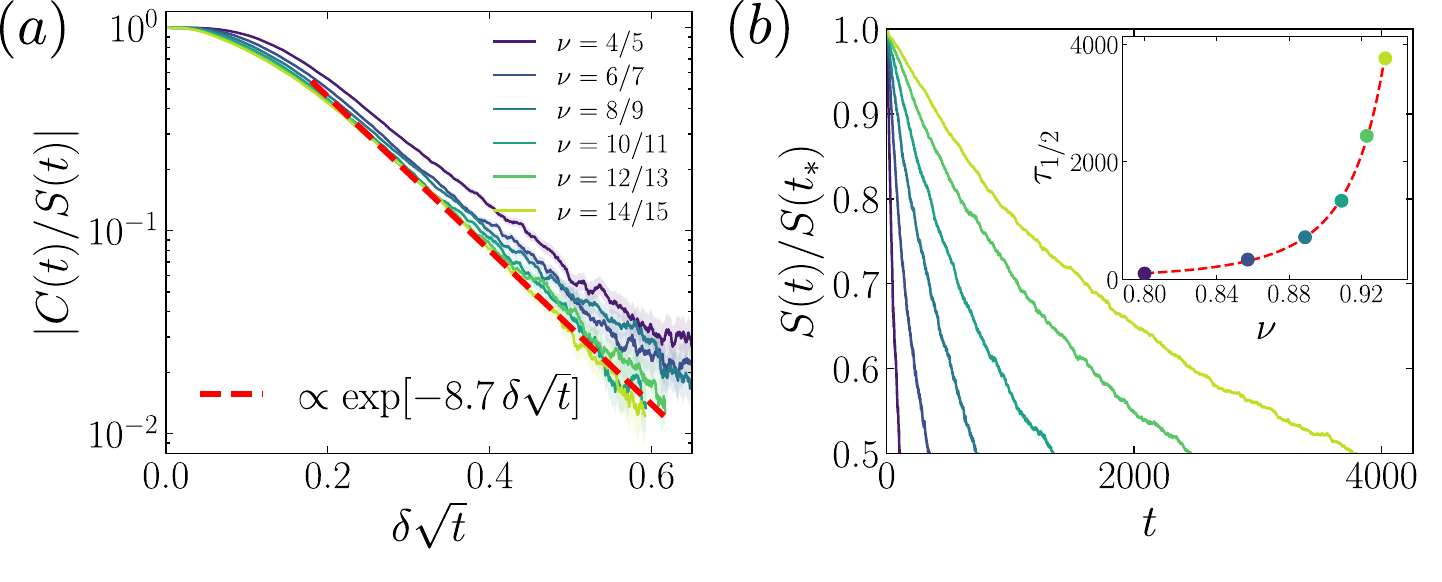}
    \caption{Time dependence of the half-chain conventional correlator $C(t)$ and string order parameter $S(t)$ under the classical, stochastic dynamics defined in the text, starting from states with perfect AFM order in squeezed space. For each filling, $\nu$, the system size is such that $24$ holes are present. (a) Ratio of $C(t)$ to $S(t)$ 
    at early times. The red dashed line is a stretched exponential fit to the $\nu=\frac{14}{15}$ data and $\delta$ is the hole density.     (b) Long-time decay of the string order, normalized by its value at an early time $t^*=5$. Colors indicate the same fillings as in (a). The inset plots the time $\tau_{1/2}$ at which $S(\tau_{1/2})=S(t_*)/2$, as a function of filling. 
    }
    \label{fig:classical-quench}
\end{figure}

To illustrate this, we
consider a quench from a product state exhibiting perfect string order, time evolving under Eq.~\eqref{eq:hamiltonian-base} with $t_{\perp,\parallel}=1$. 
For concreteness, we use as the initial state $\ket{\uparrow \downarrow 0}^{\otimes L/3}$, and present the results of tensor network simulations \cite{Zaletel2015} of the quench in Fig.~\ref{fig:quantum-quench}. Indeed, we see that, while the conventional two-point correlator washes out nearly immediately as expected in a thermalizing system, the string order parameter persists up to the numerically accessible time scale. This seems to suggest the Rydberg blockade favors the persistence of the string order.

In order to better assess the timescales over which the string order persists, 
we approximate the quantum quench with a classical stochastic process. We consider a periodic chain with two species of hardcore particles, subject to the Rydberg constraint, which hop randomly between nearest neighbors and experience spin flips at fixed and equal rates at each time step, modeling the dynamics generated by Eq.~\eqref{eq:hamiltonian-base} for $t_{\perp,\parallel}=1$.
We present the results of this simulation of this classical model in Fig.~\ref{fig:classical-quench} for quenches from initial states at fillings $\nu=p/q$; in each initial state, the unit cell consists of $p\in 2\mathbb{Z}$ particles of alternating spin followed by $q-p$ holes. 
We use periodic boundary conditions, and compute the correlators between antipodal points, averaging over all points.
We see that, indeed, while the conventional correlator decays rapidly for all fillings, the string order lasts significantly longer. 

The squeezed space picture makes manifest the role of the Rydberg blockade in this behavior.
In terms of the squeezed space variables introduced above, the initial state exhibits exact AFM order and thus uniform $\tau_i$. Let us understand the processes that lead to decay of the string and conventional order parameters. Since the string order measures magnetic order in squeezed space, $S_{ab} = -\tau_i \tau_j$ , its value can only change when a spin flips at one of its endpoints; this can happen only if that particle is flanked by two holes, due to the Rydberg blockade. The same spin flip processes change the value of the conventional correlator, but now holes hopping past $a$ and $b$ also change its value; recall $C_{ab}=S_{ab}(-1)^{N_{ab}}$. We thus expect the conventional correlator to decay \emph{faster} than the string order parameter due to these two decay processes. 

More quantitatively, in the dilute hole limit and at early times, we can treat the holes as independent diffusing particles. Indeed, before any holes come close enough to allow for a spin flip, they exactly obey the symmetric simple exclusion process \cite{Spohn1991}. During this initial period, the string order thus remains unchanged (except when a hole lands exactly on an endpoint). In contrast, the conventional correlator will start to decay due to holes diffusing through its endpoints. The decay of the correlator can therefore be estimated as the average
number of holes that pass from one side of an endpoint to the other.
Only holes within a distance $\sim\sqrt{t}$ have an appreciable probability of traversing one of the endpoints after a time $t$. 
Letting $\delta = 1-\nu$ be the hole doping,
the number of such holes is proportional to $\delta \sqrt{t}$, and thus we expect the average hole parity to decay as a stretched exponential, $e^{- A \delta \sqrt{t} }$, for some $A > 0$. 

The conventional correlator should therefore decay significantly faster than the string order parameter at any small hole doping. 
Our numerical simulations, shown in Fig.~\ref{fig:classical-quench}(a), confirm this expectation.
Now, in the limit of $\delta=0$, spin flips never occur and thus the string order is exactly preserved. We therefore expect the lifetime of the string order to diverge in the limit $\delta \to 0$. 
Indeed, as shown in Fig.~\ref{fig:classical-quench}(b), we observe the string order to persist for parametrically long times in this limit.
This demonstrates that the Rydberg blockade can be used to preserve topological correlations out of equilibrium to late times. 

\textit{Discussion---}%
In this Letter, we have proposed the $N$-leg PXYP model as a minimal setting to explore constraint-induced intertwinement between itinerant and internal degrees of freedom. In particular, we showed that the Rydberg blockade naturally induces 
(quasi-)long range squeezed space correlations, 
and hence nontrivial string order
both in and out of equilibrium. 
Importantly, this model is closely related to the physics realized in a range of experimental platforms.

Despite their simplicity, this family of models presents several natural directions for further exploration. 
We showed that the three-leg PXYP model, despite being gapless, supports algebraic string order, which is experimentally accessible in cold-atom experiments \cite{Hilker2017}. This suggests such constrained Hamiltonians are natural settings to realize novel forms of gapless topology (e.g. ``purely" gapless topology \cite{Verresen2021,Li2025}). Moreover, the persistence of the nonlocal string order at late times in the quench, when the holes are interacting, suggests the emergent hydrodynamics of this model may be interesting to study.

\textit{Acknowledgments---}%
We thank Luca Delacr\'etaz, Adam Nahum, and Gabriel Artur Weiderpass for helpful discussions.
Tensor network calculations
were performed using the TeNPy library (version 1.1.0) \cite{tenpy2024}.
We acknowledge the use of ChatGPT 5.6 and 6 for assistance in developing the numerical classical stochastic simulations and in sharpening the quantitative analysis of the stretched exponential decay and dilute particle regimes. It also pointed out the connection with Ref.~\onlinecite{Bauer2013}. The authors assume full responsibility for this work.

\bibliographystyle{apsrev4-2}
\bibliography{bib}

\begin{thebibliography}{73}%
\makeatletter
\providecommand \@ifxundefined [1]{%
 \@ifx{#1\undefined}
}%
\providecommand \@ifnum [1]{%
 \ifnum #1\expandafter \@firstoftwo
 \else \expandafter \@secondoftwo
 \fi
}%
\providecommand \@ifx [1]{%
 \ifx #1\expandafter \@firstoftwo
 \else \expandafter \@secondoftwo
 \fi
}%
\providecommand \natexlab [1]{#1}%
\providecommand \enquote  [1]{``#1''}%
\providecommand \bibnamefont  [1]{#1}%
\providecommand \bibfnamefont [1]{#1}%
\providecommand \citenamefont [1]{#1}%
\providecommand \href@noop [0]{\@secondoftwo}%
\providecommand \href [0]{\begingroup \@sanitize@url \@href}%
\providecommand \@href[1]{\@@startlink{#1}\@@href}%
\providecommand \@@href[1]{\endgroup#1\@@endlink}%
\providecommand \@sanitize@url [0]{\catcode `\\12\catcode `\$12\catcode
  `\&12\catcode `\#12\catcode `\^12\catcode `\_12\catcode `\%12\relax}%
\providecommand \@@startlink[1]{}%
\providecommand \@@endlink[0]{}%
\providecommand \url  [0]{\begingroup\@sanitize@url \@url }%
\providecommand \@url [1]{\endgroup\@href {#1}{\urlprefix }}%
\providecommand \urlprefix  [0]{URL }%
\providecommand \Eprint [0]{\href }%
\providecommand \doibase [0]{https://doi.org/}%
\providecommand \selectlanguage [0]{\@gobble}%
\providecommand \bibinfo  [0]{\@secondoftwo}%
\providecommand \bibfield  [0]{\@secondoftwo}%
\providecommand \translation [1]{[#1]}%
\providecommand \BibitemOpen [0]{}%
\providecommand \bibitemStop [0]{}%
\providecommand \bibitemNoStop [0]{.\EOS\space}%
\providecommand \EOS [0]{\spacefactor3000\relax}%
\providecommand \BibitemShut  [1]{\csname bibitem#1\endcsname}%
\let\auto@bib@innerbib\@empty
\bibitem [{\citenamefont {Rokhsar}\ and\ \citenamefont
  {Kivelson}(1988)}]{Rokhsar1988}%
  \BibitemOpen
  \bibfield  {author} {\bibinfo {author} {\bibfnamefont {D.~S.}\ \bibnamefont
  {Rokhsar}}\ and\ \bibinfo {author} {\bibfnamefont {S.~A.}\ \bibnamefont
  {Kivelson}},\ }\href {https://doi.org/10.1103/PhysRevLett.61.2376} {\bibfield
   {journal} {\bibinfo  {journal} {Phys. Rev. Lett.}\ }\textbf {\bibinfo
  {volume} {61}},\ \bibinfo {pages} {2376} (\bibinfo {year}
  {1988})}\BibitemShut {NoStop}%
\bibitem [{\citenamefont {Moessner}\ and\ \citenamefont
  {Sondhi}(2001)}]{Moessner2001}%
  \BibitemOpen
  \bibfield  {author} {\bibinfo {author} {\bibfnamefont {R.}~\bibnamefont
  {Moessner}}\ and\ \bibinfo {author} {\bibfnamefont {S.~L.}\ \bibnamefont
  {Sondhi}},\ }\href {https://doi.org/10.1103/PhysRevLett.86.1881} {\bibfield
  {journal} {\bibinfo  {journal} {Phys. Rev. Lett.}\ }\textbf {\bibinfo
  {volume} {86}},\ \bibinfo {pages} {1881} (\bibinfo {year} {2001})},\ \Eprint
  {https://arxiv.org/abs/cond-mat/0007378} {arXiv:cond-mat/0007378}
  \BibitemShut {NoStop}%
\bibitem [{\citenamefont {Misguich}\ \emph {et~al.}(2002)\citenamefont
  {Misguich}, \citenamefont {Serban},\ and\ \citenamefont
  {Pasquier}}]{Misguich2002}%
  \BibitemOpen
  \bibfield  {author} {\bibinfo {author} {\bibfnamefont {G.}~\bibnamefont
  {Misguich}}, \bibinfo {author} {\bibfnamefont {D.}~\bibnamefont {Serban}},\
  and\ \bibinfo {author} {\bibfnamefont {V.}~\bibnamefont {Pasquier}},\ }\href
  {https://doi.org/10.1103/PhysRevLett.89.137202} {\bibfield  {journal}
  {\bibinfo  {journal} {Phys. Rev. Lett.}\ }\textbf {\bibinfo {volume} {89}},\
  \bibinfo {pages} {137202} (\bibinfo {year} {2002})},\ \Eprint
  {https://arxiv.org/abs/cond-mat/0204428} {arXiv:cond-mat/0204428}
  \BibitemShut {NoStop}%
\bibitem [{\citenamefont {Fradkin}(2013)}]{Fradkin2013}%
  \BibitemOpen
  \bibfield  {author} {\bibinfo {author} {\bibfnamefont {E.}~\bibnamefont
  {Fradkin}},\ }\href {https://doi.org/10.1017/CBO9781139015509} {\emph
  {\bibinfo {title} {Field Theories of Condensed Matter Physics}}},\ \bibinfo
  {edition} {2nd}\ ed.\ (\bibinfo  {publisher} {Cambridge University Press},\
  \bibinfo {address} {Cambridge},\ \bibinfo {year} {2013})\BibitemShut
  {NoStop}%
\bibitem [{\citenamefont {Turner}\ \emph
  {et~al.}(2018{\natexlab{a}})\citenamefont {Turner}, \citenamefont
  {Michailidis}, \citenamefont {Abanin}, \citenamefont {Serbyn},\ and\
  \citenamefont {Papi{\'c}}}]{Turner2018a}%
  \BibitemOpen
  \bibfield  {author} {\bibinfo {author} {\bibfnamefont {C.~J.}\ \bibnamefont
  {Turner}}, \bibinfo {author} {\bibfnamefont {A.~A.}\ \bibnamefont
  {Michailidis}}, \bibinfo {author} {\bibfnamefont {D.~A.}\ \bibnamefont
  {Abanin}}, \bibinfo {author} {\bibfnamefont {M.}~\bibnamefont {Serbyn}},\
  and\ \bibinfo {author} {\bibfnamefont {Z.}~\bibnamefont {Papi{\'c}}},\ }\href
  {https://doi.org/10.1038/s41567-018-0137-5} {\bibfield  {journal} {\bibinfo
  {journal} {Nat. Phys.}\ }\textbf {\bibinfo {volume} {14}},\ \bibinfo {pages}
  {745} (\bibinfo {year} {2018}{\natexlab{a}})},\ \Eprint
  {https://arxiv.org/abs/1711.03528} {arXiv:1711.03528} \BibitemShut {NoStop}%
\bibitem [{\citenamefont {Turner}\ \emph
  {et~al.}(2018{\natexlab{b}})\citenamefont {Turner}, \citenamefont
  {Michailidis}, \citenamefont {Abanin}, \citenamefont {Serbyn},\ and\
  \citenamefont {Papi{\'c}}}]{Turner2018b}%
  \BibitemOpen
  \bibfield  {author} {\bibinfo {author} {\bibfnamefont {C.~J.}\ \bibnamefont
  {Turner}}, \bibinfo {author} {\bibfnamefont {A.~A.}\ \bibnamefont
  {Michailidis}}, \bibinfo {author} {\bibfnamefont {D.~A.}\ \bibnamefont
  {Abanin}}, \bibinfo {author} {\bibfnamefont {M.}~\bibnamefont {Serbyn}},\
  and\ \bibinfo {author} {\bibfnamefont {Z.}~\bibnamefont {Papi{\'c}}},\ }\href
  {https://doi.org/10.1103/PhysRevB.98.155134} {\bibfield  {journal} {\bibinfo
  {journal} {Phys. Rev. B}\ }\textbf {\bibinfo {volume} {98}},\ \bibinfo
  {pages} {155134} (\bibinfo {year} {2018}{\natexlab{b}})},\ \Eprint
  {https://arxiv.org/abs/1806.10933} {arXiv:1806.10933} \BibitemShut {NoStop}%
\bibitem [{\citenamefont {Sala}\ \emph {et~al.}(2020)\citenamefont {Sala},
  \citenamefont {Rakovszky}, \citenamefont {Verresen}, \citenamefont {Knap},\
  and\ \citenamefont {Pollmann}}]{Sala2020}%
  \BibitemOpen
  \bibfield  {author} {\bibinfo {author} {\bibfnamefont {P.}~\bibnamefont
  {Sala}}, \bibinfo {author} {\bibfnamefont {T.}~\bibnamefont {Rakovszky}},
  \bibinfo {author} {\bibfnamefont {R.}~\bibnamefont {Verresen}}, \bibinfo
  {author} {\bibfnamefont {M.}~\bibnamefont {Knap}},\ and\ \bibinfo {author}
  {\bibfnamefont {F.}~\bibnamefont {Pollmann}},\ }\href
  {https://doi.org/10.1103/PhysRevX.10.011047} {\bibfield  {journal} {\bibinfo
  {journal} {Phys. Rev. X}\ }\textbf {\bibinfo {volume} {10}},\ \bibinfo
  {pages} {011047} (\bibinfo {year} {2020})},\ \Eprint
  {https://arxiv.org/abs/1904.04266} {arXiv:1904.04266} \BibitemShut {NoStop}%
\bibitem [{\citenamefont {Khemani}\ \emph {et~al.}(2020)\citenamefont
  {Khemani}, \citenamefont {Hermele},\ and\ \citenamefont
  {Nandkishore}}]{Khemani2020}%
  \BibitemOpen
  \bibfield  {author} {\bibinfo {author} {\bibfnamefont {V.}~\bibnamefont
  {Khemani}}, \bibinfo {author} {\bibfnamefont {M.}~\bibnamefont {Hermele}},\
  and\ \bibinfo {author} {\bibfnamefont {R.}~\bibnamefont {Nandkishore}},\
  }\href {https://doi.org/10.1103/PhysRevB.101.174204} {\bibfield  {journal}
  {\bibinfo  {journal} {Phys. Rev. B}\ }\textbf {\bibinfo {volume} {101}},\
  \bibinfo {pages} {174204} (\bibinfo {year} {2020})},\ \Eprint
  {https://arxiv.org/abs/1904.04815} {arXiv:1904.04815} \BibitemShut {NoStop}%
\bibitem [{\citenamefont {Serbyn}\ \emph {et~al.}(2021)\citenamefont {Serbyn},
  \citenamefont {Abanin},\ and\ \citenamefont {Papi{\'c}}}]{Serbyn2021}%
  \BibitemOpen
  \bibfield  {author} {\bibinfo {author} {\bibfnamefont {M.}~\bibnamefont
  {Serbyn}}, \bibinfo {author} {\bibfnamefont {D.~A.}\ \bibnamefont {Abanin}},\
  and\ \bibinfo {author} {\bibfnamefont {Z.}~\bibnamefont {Papi{\'c}}},\ }\href
  {https://doi.org/10.1038/s41567-021-01230-2} {\bibfield  {journal} {\bibinfo
  {journal} {Nat. Phys.}\ }\textbf {\bibinfo {volume} {17}},\ \bibinfo {pages}
  {675} (\bibinfo {year} {2021})},\ \Eprint {https://arxiv.org/abs/2011.09486}
  {arXiv:2011.09486} \BibitemShut {NoStop}%
\bibitem [{\citenamefont {Moudgalya}\ \emph {et~al.}(2022)\citenamefont
  {Moudgalya}, \citenamefont {Bernevig},\ and\ \citenamefont
  {Regnault}}]{Moudgalya2022a}%
  \BibitemOpen
  \bibfield  {author} {\bibinfo {author} {\bibfnamefont {S.}~\bibnamefont
  {Moudgalya}}, \bibinfo {author} {\bibfnamefont {B.~A.}\ \bibnamefont
  {Bernevig}},\ and\ \bibinfo {author} {\bibfnamefont {N.}~\bibnamefont
  {Regnault}},\ }\href {https://doi.org/10.1088/1361-6633/ac73a0} {\bibfield
  {journal} {\bibinfo  {journal} {Rep. Prog. Phys.}\ }\textbf {\bibinfo
  {volume} {85}},\ \bibinfo {pages} {086501} (\bibinfo {year} {2022})},\
  \Eprint {https://arxiv.org/abs/2109.00548} {arXiv:2109.00548} \BibitemShut
  {NoStop}%
\bibitem [{\citenamefont {Browaeys}\ and\ \citenamefont
  {Lahaye}(2020)}]{Browaeys2020}%
  \BibitemOpen
  \bibfield  {author} {\bibinfo {author} {\bibfnamefont {A.}~\bibnamefont
  {Browaeys}}\ and\ \bibinfo {author} {\bibfnamefont {T.}~\bibnamefont
  {Lahaye}},\ }\href {https://doi.org/10.1038/s41567-019-0733-z} {\bibfield
  {journal} {\bibinfo  {journal} {Nature Physics}\ }\textbf {\bibinfo {volume}
  {16}},\ \bibinfo {pages} {132–142} (\bibinfo {year} {2020})}\BibitemShut
  {NoStop}%
\bibitem [{\citenamefont {Fendley}\ \emph {et~al.}(2004)\citenamefont
  {Fendley}, \citenamefont {Sengupta},\ and\ \citenamefont
  {Sachdev}}]{Fendley2004}%
  \BibitemOpen
  \bibfield  {author} {\bibinfo {author} {\bibfnamefont {P.}~\bibnamefont
  {Fendley}}, \bibinfo {author} {\bibfnamefont {K.}~\bibnamefont {Sengupta}},\
  and\ \bibinfo {author} {\bibfnamefont {S.}~\bibnamefont {Sachdev}},\
  }\bibfield  {journal} {\bibinfo  {journal} {Physical Review B}\ }\textbf
  {\bibinfo {volume} {69}},\ \href {https://doi.org/10.1103/physrevb.69.075106}
  {10.1103/physrevb.69.075106} (\bibinfo {year} {2004})\BibitemShut {NoStop}%
\bibitem [{\citenamefont {Lesanovsky}\ and\ \citenamefont
  {Katsura}(2012)}]{Lesanovsky2012}%
  \BibitemOpen
  \bibfield  {author} {\bibinfo {author} {\bibfnamefont {I.}~\bibnamefont
  {Lesanovsky}}\ and\ \bibinfo {author} {\bibfnamefont {H.}~\bibnamefont
  {Katsura}},\ }\bibfield  {journal} {\bibinfo  {journal} {Physical Review A}\
  }\textbf {\bibinfo {volume} {86}},\ \href
  {https://doi.org/10.1103/physreva.86.041601} {10.1103/physreva.86.041601}
  (\bibinfo {year} {2012})\BibitemShut {NoStop}%
\bibitem [{\citenamefont {Bernien}\ \emph {et~al.}(2017)\citenamefont
  {Bernien}, \citenamefont {Schwartz}, \citenamefont {Keesling}, \citenamefont
  {Levine}, \citenamefont {Omran}, \citenamefont {Pichler}, \citenamefont
  {Choi}, \citenamefont {Zibrov}, \citenamefont {Endres}, \citenamefont
  {Greiner}, \citenamefont {Vuleti{\'c}},\ and\ \citenamefont
  {Lukin}}]{Bernien2017}%
  \BibitemOpen
  \bibfield  {author} {\bibinfo {author} {\bibfnamefont {H.}~\bibnamefont
  {Bernien}}, \bibinfo {author} {\bibfnamefont {S.}~\bibnamefont {Schwartz}},
  \bibinfo {author} {\bibfnamefont {A.}~\bibnamefont {Keesling}}, \bibinfo
  {author} {\bibfnamefont {H.}~\bibnamefont {Levine}}, \bibinfo {author}
  {\bibfnamefont {A.}~\bibnamefont {Omran}}, \bibinfo {author} {\bibfnamefont
  {H.}~\bibnamefont {Pichler}}, \bibinfo {author} {\bibfnamefont
  {S.}~\bibnamefont {Choi}}, \bibinfo {author} {\bibfnamefont {A.~S.}\
  \bibnamefont {Zibrov}}, \bibinfo {author} {\bibfnamefont {M.}~\bibnamefont
  {Endres}}, \bibinfo {author} {\bibfnamefont {M.}~\bibnamefont {Greiner}},
  \bibinfo {author} {\bibfnamefont {V.}~\bibnamefont {Vuleti{\'c}}},\ and\
  \bibinfo {author} {\bibfnamefont {M.~D.}\ \bibnamefont {Lukin}},\ }\href
  {https://doi.org/10.1038/nature24622} {\bibfield  {journal} {\bibinfo
  {journal} {Nature}\ }\textbf {\bibinfo {volume} {551}},\ \bibinfo {pages}
  {579} (\bibinfo {year} {2017})},\ \Eprint {https://arxiv.org/abs/1707.04344}
  {arXiv:1707.04344} \BibitemShut {NoStop}%
\bibitem [{\citenamefont {Glaetzle}\ \emph {et~al.}(2014)\citenamefont
  {Glaetzle}, \citenamefont {Dalmonte}, \citenamefont {Nath}, \citenamefont
  {Rousochatzakis}, \citenamefont {Moessner},\ and\ \citenamefont
  {Zoller}}]{Glaetzle2014}%
  \BibitemOpen
  \bibfield  {author} {\bibinfo {author} {\bibfnamefont {A.~W.}\ \bibnamefont
  {Glaetzle}}, \bibinfo {author} {\bibfnamefont {M.}~\bibnamefont {Dalmonte}},
  \bibinfo {author} {\bibfnamefont {R.}~\bibnamefont {Nath}}, \bibinfo {author}
  {\bibfnamefont {I.}~\bibnamefont {Rousochatzakis}}, \bibinfo {author}
  {\bibfnamefont {R.}~\bibnamefont {Moessner}},\ and\ \bibinfo {author}
  {\bibfnamefont {P.}~\bibnamefont {Zoller}},\ }\href
  {https://doi.org/10.1103/PhysRevX.4.041037} {\bibfield  {journal} {\bibinfo
  {journal} {Phys. Rev. X}\ }\textbf {\bibinfo {volume} {4}},\ \bibinfo {pages}
  {041037} (\bibinfo {year} {2014})},\ \Eprint
  {https://arxiv.org/abs/1404.5326} {arXiv:1404.5326} \BibitemShut {NoStop}%
\bibitem [{\citenamefont {Celi}\ \emph {et~al.}(2020)\citenamefont {Celi},
  \citenamefont {Vermersch}, \citenamefont {Viyuela}, \citenamefont {Pichler},
  \citenamefont {Lukin},\ and\ \citenamefont {Zoller}}]{Celi2020}%
  \BibitemOpen
  \bibfield  {author} {\bibinfo {author} {\bibfnamefont {A.}~\bibnamefont
  {Celi}}, \bibinfo {author} {\bibfnamefont {B.}~\bibnamefont {Vermersch}},
  \bibinfo {author} {\bibfnamefont {O.}~\bibnamefont {Viyuela}}, \bibinfo
  {author} {\bibfnamefont {H.}~\bibnamefont {Pichler}}, \bibinfo {author}
  {\bibfnamefont {M.~D.}\ \bibnamefont {Lukin}},\ and\ \bibinfo {author}
  {\bibfnamefont {P.}~\bibnamefont {Zoller}},\ }\href
  {https://doi.org/10.1103/PhysRevX.10.021057} {\bibfield  {journal} {\bibinfo
  {journal} {Phys. Rev. X}\ }\textbf {\bibinfo {volume} {10}},\ \bibinfo
  {pages} {021057} (\bibinfo {year} {2020})},\ \Eprint
  {https://arxiv.org/abs/1907.03311} {arXiv:1907.03311} \BibitemShut {NoStop}%
\bibitem [{\citenamefont {Surace}\ \emph {et~al.}(2020)\citenamefont {Surace},
  \citenamefont {Mazza}, \citenamefont {Giudici}, \citenamefont {Lerose},
  \citenamefont {Gambassi},\ and\ \citenamefont {Dalmonte}}]{Surace2020}%
  \BibitemOpen
  \bibfield  {author} {\bibinfo {author} {\bibfnamefont {F.~M.}\ \bibnamefont
  {Surace}}, \bibinfo {author} {\bibfnamefont {P.~P.}\ \bibnamefont {Mazza}},
  \bibinfo {author} {\bibfnamefont {G.}~\bibnamefont {Giudici}}, \bibinfo
  {author} {\bibfnamefont {A.}~\bibnamefont {Lerose}}, \bibinfo {author}
  {\bibfnamefont {A.}~\bibnamefont {Gambassi}},\ and\ \bibinfo {author}
  {\bibfnamefont {M.}~\bibnamefont {Dalmonte}},\ }\href
  {https://doi.org/10.1103/PhysRevX.10.021041} {\bibfield  {journal} {\bibinfo
  {journal} {Phys. Rev. X}\ }\textbf {\bibinfo {volume} {10}},\ \bibinfo
  {pages} {021041} (\bibinfo {year} {2020})},\ \Eprint
  {https://arxiv.org/abs/1902.09551} {arXiv:1902.09551} \BibitemShut {NoStop}%
\bibitem [{\citenamefont {Verresen}\ \emph
  {et~al.}(2021{\natexlab{a}})\citenamefont {Verresen}, \citenamefont {Lukin},\
  and\ \citenamefont {Vishwanath}}]{Verresen2021a}%
  \BibitemOpen
  \bibfield  {author} {\bibinfo {author} {\bibfnamefont {R.}~\bibnamefont
  {Verresen}}, \bibinfo {author} {\bibfnamefont {M.~D.}\ \bibnamefont
  {Lukin}},\ and\ \bibinfo {author} {\bibfnamefont {A.}~\bibnamefont
  {Vishwanath}},\ }\href {https://doi.org/10.1103/PhysRevX.11.031005}
  {\bibfield  {journal} {\bibinfo  {journal} {Phys. Rev. X}\ }\textbf {\bibinfo
  {volume} {11}},\ \bibinfo {pages} {031005} (\bibinfo {year}
  {2021}{\natexlab{a}})}\BibitemShut {NoStop}%
\bibitem [{\citenamefont {Samajdar}\ \emph {et~al.}(2021)\citenamefont
  {Samajdar}, \citenamefont {Ho}, \citenamefont {Pichler}, \citenamefont
  {Lukin},\ and\ \citenamefont {Sachdev}}]{Samajdar2021}%
  \BibitemOpen
  \bibfield  {author} {\bibinfo {author} {\bibfnamefont {R.}~\bibnamefont
  {Samajdar}}, \bibinfo {author} {\bibfnamefont {W.~W.}\ \bibnamefont {Ho}},
  \bibinfo {author} {\bibfnamefont {H.}~\bibnamefont {Pichler}}, \bibinfo
  {author} {\bibfnamefont {M.~D.}\ \bibnamefont {Lukin}},\ and\ \bibinfo
  {author} {\bibfnamefont {S.}~\bibnamefont {Sachdev}},\ }\href
  {https://doi.org/10.1073/pnas.2015785118} {\bibfield  {journal} {\bibinfo
  {journal} {Proc. Natl. Acad. Sci. U.S.A.}\ }\textbf {\bibinfo {volume}
  {118}},\ \bibinfo {pages} {e2015785118} (\bibinfo {year} {2021})},\ \Eprint
  {https://arxiv.org/abs/2011.12295} {arXiv:2011.12295} \BibitemShut {NoStop}%
\bibitem [{\citenamefont {Giudici}\ \emph {et~al.}(2022)\citenamefont
  {Giudici}, \citenamefont {Lukin},\ and\ \citenamefont
  {Pichler}}]{Giudici2022}%
  \BibitemOpen
  \bibfield  {author} {\bibinfo {author} {\bibfnamefont {G.}~\bibnamefont
  {Giudici}}, \bibinfo {author} {\bibfnamefont {M.~D.}\ \bibnamefont {Lukin}},\
  and\ \bibinfo {author} {\bibfnamefont {H.}~\bibnamefont {Pichler}},\
  }\bibfield  {journal} {\bibinfo  {journal} {Physical Review Letters}\
  }\textbf {\bibinfo {volume} {129}},\ \href
  {https://doi.org/10.1103/physrevlett.129.090401}
  {10.1103/physrevlett.129.090401} (\bibinfo {year} {2022})\BibitemShut
  {NoStop}%
\bibitem [{\citenamefont {Tarabunga}\ \emph {et~al.}(2022)\citenamefont
  {Tarabunga}, \citenamefont {Surace}, \citenamefont {Andreoni}, \citenamefont
  {Angelone},\ and\ \citenamefont {Dalmonte}}]{Tarabunga2022}%
  \BibitemOpen
  \bibfield  {author} {\bibinfo {author} {\bibfnamefont {P.~S.}\ \bibnamefont
  {Tarabunga}}, \bibinfo {author} {\bibfnamefont {F.~M.}\ \bibnamefont
  {Surace}}, \bibinfo {author} {\bibfnamefont {R.}~\bibnamefont {Andreoni}},
  \bibinfo {author} {\bibfnamefont {A.}~\bibnamefont {Angelone}},\ and\
  \bibinfo {author} {\bibfnamefont {M.}~\bibnamefont {Dalmonte}},\ }\href
  {https://doi.org/10.1103/PhysRevLett.129.195301} {\bibfield  {journal}
  {\bibinfo  {journal} {Phys. Rev. Lett.}\ }\textbf {\bibinfo {volume} {129}},\
  \bibinfo {pages} {195301} (\bibinfo {year} {2022})},\ \Eprint
  {https://arxiv.org/abs/2205.13000} {arXiv:2205.13000} \BibitemShut {NoStop}%
\bibitem [{\citenamefont {Verresen}\ and\ \citenamefont
  {Vishwanath}(2022)}]{Verresen2022}%
  \BibitemOpen
  \bibfield  {author} {\bibinfo {author} {\bibfnamefont {R.}~\bibnamefont
  {Verresen}}\ and\ \bibinfo {author} {\bibfnamefont {A.}~\bibnamefont
  {Vishwanath}},\ }\href {https://doi.org/10.1103/PhysRevX.12.041029}
  {\bibfield  {journal} {\bibinfo  {journal} {Phys. Rev. X}\ }\textbf {\bibinfo
  {volume} {12}},\ \bibinfo {pages} {041029} (\bibinfo {year} {2022})},\
  \Eprint {https://arxiv.org/abs/2205.15302} {arXiv:2205.15302} \BibitemShut
  {NoStop}%
\bibitem [{\citenamefont {Samajdar}\ \emph {et~al.}(2023)\citenamefont
  {Samajdar}, \citenamefont {Joshi}, \citenamefont {Teng},\ and\ \citenamefont
  {Sachdev}}]{Samajdar2023}%
  \BibitemOpen
  \bibfield  {author} {\bibinfo {author} {\bibfnamefont {R.}~\bibnamefont
  {Samajdar}}, \bibinfo {author} {\bibfnamefont {D.~G.}\ \bibnamefont {Joshi}},
  \bibinfo {author} {\bibfnamefont {Y.}~\bibnamefont {Teng}},\ and\ \bibinfo
  {author} {\bibfnamefont {S.}~\bibnamefont {Sachdev}},\ }\href
  {https://doi.org/10.1103/PhysRevLett.130.043601} {\bibfield  {journal}
  {\bibinfo  {journal} {Phys. Rev. Lett.}\ }\textbf {\bibinfo {volume} {130}},\
  \bibinfo {pages} {043601} (\bibinfo {year} {2023})}\BibitemShut {NoStop}%
\bibitem [{\citenamefont {Sahay}\ \emph {et~al.}(2023)\citenamefont {Sahay},
  \citenamefont {Vishwanath},\ and\ \citenamefont {Verresen}}]{Sahay2023}%
  \BibitemOpen
  \bibfield  {author} {\bibinfo {author} {\bibfnamefont {R.}~\bibnamefont
  {Sahay}}, \bibinfo {author} {\bibfnamefont {A.}~\bibnamefont {Vishwanath}},\
  and\ \bibinfo {author} {\bibfnamefont {R.}~\bibnamefont {Verresen}},\ }\href
  {https://arxiv.org/abs/2211.01381} {\bibinfo {title} {Quantum spin puddles
  and lakes: Nisq-era spin liquids from non-equilibrium dynamics}} (\bibinfo
  {year} {2023}),\ \Eprint {https://arxiv.org/abs/2211.01381} {arXiv:2211.01381
  [cond-mat.str-el]} \BibitemShut {NoStop}%
\bibitem [{\citenamefont {Shah}\ \emph {et~al.}(2025)\citenamefont {Shah},
  \citenamefont {Nambiar}, \citenamefont {Gorshkov},\ and\ \citenamefont
  {Galitski}}]{Shah2025}%
  \BibitemOpen
  \bibfield  {author} {\bibinfo {author} {\bibfnamefont {J.}~\bibnamefont
  {Shah}}, \bibinfo {author} {\bibfnamefont {G.}~\bibnamefont {Nambiar}},
  \bibinfo {author} {\bibfnamefont {A.~V.}\ \bibnamefont {Gorshkov}},\ and\
  \bibinfo {author} {\bibfnamefont {V.}~\bibnamefont {Galitski}},\ }\href
  {https://doi.org/10.1103/PhysRevX.15.011025} {\bibfield  {journal} {\bibinfo
  {journal} {Phys. Rev. X}\ }\textbf {\bibinfo {volume} {15}},\ \bibinfo
  {pages} {011025} (\bibinfo {year} {2025})},\ \Eprint
  {https://arxiv.org/abs/2301.04657} {arXiv:2301.04657} \BibitemShut {NoStop}%
\bibitem [{\citenamefont {Zeng}\ \emph {et~al.}(2025)\citenamefont {Zeng},
  \citenamefont {Giudici},\ and\ \citenamefont {Pichler}}]{Zeng2025}%
  \BibitemOpen
  \bibfield  {author} {\bibinfo {author} {\bibfnamefont {Z.}~\bibnamefont
  {Zeng}}, \bibinfo {author} {\bibfnamefont {G.}~\bibnamefont {Giudici}},\ and\
  \bibinfo {author} {\bibfnamefont {H.}~\bibnamefont {Pichler}},\ }\href
  {https://doi.org/10.1103/PhysRevResearch.7.L012006} {\bibfield  {journal}
  {\bibinfo  {journal} {Phys. Rev. Research}\ }\textbf {\bibinfo {volume}
  {7}},\ \bibinfo {pages} {L012006} (\bibinfo {year} {2025})},\ \Eprint
  {https://arxiv.org/abs/2402.10651} {arXiv:2402.10651} \BibitemShut {NoStop}%
\bibitem [{\citenamefont {Semeghini}\ \emph {et~al.}(2021)\citenamefont
  {Semeghini}, \citenamefont {Levine}, \citenamefont {Keesling}, \citenamefont
  {Ebadi}, \citenamefont {Wang}, \citenamefont {Bluvstein}, \citenamefont
  {Verresen}, \citenamefont {Pichler}, \citenamefont {Kalinowski},
  \citenamefont {Samajdar}, \citenamefont {Omran}, \citenamefont {Sachdev},
  \citenamefont {Vishwanath}, \citenamefont {Greiner}, \citenamefont
  {Vuletić},\ and\ \citenamefont {Lukin}}]{Semeghini2021}%
  \BibitemOpen
  \bibfield  {author} {\bibinfo {author} {\bibfnamefont {G.}~\bibnamefont
  {Semeghini}}, \bibinfo {author} {\bibfnamefont {H.}~\bibnamefont {Levine}},
  \bibinfo {author} {\bibfnamefont {A.}~\bibnamefont {Keesling}}, \bibinfo
  {author} {\bibfnamefont {S.}~\bibnamefont {Ebadi}}, \bibinfo {author}
  {\bibfnamefont {T.~T.}\ \bibnamefont {Wang}}, \bibinfo {author}
  {\bibfnamefont {D.}~\bibnamefont {Bluvstein}}, \bibinfo {author}
  {\bibfnamefont {R.}~\bibnamefont {Verresen}}, \bibinfo {author}
  {\bibfnamefont {H.}~\bibnamefont {Pichler}}, \bibinfo {author} {\bibfnamefont
  {M.}~\bibnamefont {Kalinowski}}, \bibinfo {author} {\bibfnamefont
  {R.}~\bibnamefont {Samajdar}}, \bibinfo {author} {\bibfnamefont
  {A.}~\bibnamefont {Omran}}, \bibinfo {author} {\bibfnamefont
  {S.}~\bibnamefont {Sachdev}}, \bibinfo {author} {\bibfnamefont
  {A.}~\bibnamefont {Vishwanath}}, \bibinfo {author} {\bibfnamefont
  {M.}~\bibnamefont {Greiner}}, \bibinfo {author} {\bibfnamefont
  {V.}~\bibnamefont {Vuletić}},\ and\ \bibinfo {author} {\bibfnamefont
  {M.~D.}\ \bibnamefont {Lukin}},\ }\href
  {https://doi.org/10.1126/science.abi8794} {\bibfield  {journal} {\bibinfo
  {journal} {Science}\ }\textbf {\bibinfo {volume} {374}},\ \bibinfo {pages}
  {1242–1247} (\bibinfo {year} {2021})}\BibitemShut {NoStop}%
\bibitem [{\citenamefont {Geim}\ \emph {et~al.}(2026)\citenamefont {Geim},
  \citenamefont {Koyluoglu}, \citenamefont {Evered}, \citenamefont {Sahay},
  \citenamefont {Li}, \citenamefont {Xu}, \citenamefont {Bluvstein},
  \citenamefont {Gjonbalaj}, \citenamefont {Maskara}, \citenamefont
  {Kalinowski}, \citenamefont {Manovitz}, \citenamefont {Verresen},
  \citenamefont {Yelin}, \citenamefont {Feldmeier}, \citenamefont {Greiner},
  \citenamefont {Vuletic},\ and\ \citenamefont {Lukin}}]{Geim2026}%
  \BibitemOpen
  \bibfield  {author} {\bibinfo {author} {\bibfnamefont {A.~A.}\ \bibnamefont
  {Geim}}, \bibinfo {author} {\bibfnamefont {N.~U.}\ \bibnamefont {Koyluoglu}},
  \bibinfo {author} {\bibfnamefont {S.~J.}\ \bibnamefont {Evered}}, \bibinfo
  {author} {\bibfnamefont {R.}~\bibnamefont {Sahay}}, \bibinfo {author}
  {\bibfnamefont {S.~H.}\ \bibnamefont {Li}}, \bibinfo {author} {\bibfnamefont
  {M.}~\bibnamefont {Xu}}, \bibinfo {author} {\bibfnamefont {D.}~\bibnamefont
  {Bluvstein}}, \bibinfo {author} {\bibfnamefont {N.~O.}\ \bibnamefont
  {Gjonbalaj}}, \bibinfo {author} {\bibfnamefont {N.}~\bibnamefont {Maskara}},
  \bibinfo {author} {\bibfnamefont {M.}~\bibnamefont {Kalinowski}}, \bibinfo
  {author} {\bibfnamefont {T.}~\bibnamefont {Manovitz}}, \bibinfo {author}
  {\bibfnamefont {R.}~\bibnamefont {Verresen}}, \bibinfo {author}
  {\bibfnamefont {S.~F.}\ \bibnamefont {Yelin}}, \bibinfo {author}
  {\bibfnamefont {J.}~\bibnamefont {Feldmeier}}, \bibinfo {author}
  {\bibfnamefont {M.}~\bibnamefont {Greiner}}, \bibinfo {author} {\bibfnamefont
  {V.}~\bibnamefont {Vuletic}},\ and\ \bibinfo {author} {\bibfnamefont {M.~D.}\
  \bibnamefont {Lukin}},\ }\href {https://arxiv.org/abs/2602.18555} {\bibinfo
  {title} {Engineering quantum criticality and dynamics on an analog-digital
  simulator}} (\bibinfo {year} {2026}),\ \Eprint
  {https://arxiv.org/abs/2602.18555} {arXiv:2602.18555 [quant-ph]} \BibitemShut
  {NoStop}%
\bibitem [{\citenamefont {Karch}\ \emph {et~al.}(2026)\citenamefont {Karch},
  \citenamefont {Will}, \citenamefont {Rodriguez}, \citenamefont {Liebster},
  \citenamefont {Huh}, \citenamefont {Knap}, \citenamefont {Pollmann},
  \citenamefont {Kuhlenkamp}, \citenamefont {Bloch},\ and\ \citenamefont
  {Aidelsburger}}]{Karch2026}%
  \BibitemOpen
  \bibfield  {author} {\bibinfo {author} {\bibfnamefont {S.}~\bibnamefont
  {Karch}}, \bibinfo {author} {\bibfnamefont {M.}~\bibnamefont {Will}},
  \bibinfo {author} {\bibfnamefont {I.~P.}\ \bibnamefont {Rodriguez}}, \bibinfo
  {author} {\bibfnamefont {N.}~\bibnamefont {Liebster}}, \bibinfo {author}
  {\bibfnamefont {S.}~\bibnamefont {Huh}}, \bibinfo {author} {\bibfnamefont
  {M.}~\bibnamefont {Knap}}, \bibinfo {author} {\bibfnamefont {F.}~\bibnamefont
  {Pollmann}}, \bibinfo {author} {\bibfnamefont {C.}~\bibnamefont
  {Kuhlenkamp}}, \bibinfo {author} {\bibfnamefont {I.}~\bibnamefont {Bloch}},\
  and\ \bibinfo {author} {\bibfnamefont {M.}~\bibnamefont {Aidelsburger}},\
  }\href {https://arxiv.org/abs/2604.24744} {\bibinfo {title} {Dynamical
  preparation of u(1) quantum spin liquids in an analogue quantum simulator}}
  (\bibinfo {year} {2026}),\ \Eprint {https://arxiv.org/abs/2604.24744}
  {arXiv:2604.24744 [cond-mat.quant-gas]} \BibitemShut {NoStop}%
\bibitem [{\citenamefont {Bornet}\ \emph {et~al.}(2026)\citenamefont {Bornet},
  \citenamefont {Bintz}, \citenamefont {Chen}, \citenamefont {Emperauger},
  \citenamefont {Qiao}, \citenamefont {Martin}, \citenamefont {Barredo},
  \citenamefont {Chatterjee}, \citenamefont {Liu}, \citenamefont {Lahaye},
  \citenamefont {Zaletel}, \citenamefont {Yao},\ and\ \citenamefont
  {Browaeys}}]{Bornet2026}%
  \BibitemOpen
  \bibfield  {author} {\bibinfo {author} {\bibfnamefont {G.}~\bibnamefont
  {Bornet}}, \bibinfo {author} {\bibfnamefont {M.}~\bibnamefont {Bintz}},
  \bibinfo {author} {\bibfnamefont {C.}~\bibnamefont {Chen}}, \bibinfo {author}
  {\bibfnamefont {G.}~\bibnamefont {Emperauger}}, \bibinfo {author}
  {\bibfnamefont {M.}~\bibnamefont {Qiao}}, \bibinfo {author} {\bibfnamefont
  {R.}~\bibnamefont {Martin}}, \bibinfo {author} {\bibfnamefont
  {D.}~\bibnamefont {Barredo}}, \bibinfo {author} {\bibfnamefont
  {S.}~\bibnamefont {Chatterjee}}, \bibinfo {author} {\bibfnamefont {V.~S.}\
  \bibnamefont {Liu}}, \bibinfo {author} {\bibfnamefont {T.}~\bibnamefont
  {Lahaye}}, \bibinfo {author} {\bibfnamefont {M.~P.}\ \bibnamefont {Zaletel}},
  \bibinfo {author} {\bibfnamefont {N.~Y.}\ \bibnamefont {Yao}},\ and\ \bibinfo
  {author} {\bibfnamefont {A.}~\bibnamefont {Browaeys}},\ }\href
  {https://arxiv.org/abs/2602.14323} {\bibinfo {title} {Dirac spin liquid
  candidate in a rydberg quantum simulator}} (\bibinfo {year} {2026}),\ \Eprint
  {https://arxiv.org/abs/2602.14323} {arXiv:2602.14323 [cond-mat.quant-gas]}
  \BibitemShut {NoStop}%
\bibitem [{\citenamefont {Singh}\ \emph {et~al.}(2021)\citenamefont {Singh},
  \citenamefont {Ware}, \citenamefont {Vasseur},\ and\ \citenamefont
  {Friedman}}]{Singh21}%
  \BibitemOpen
  \bibfield  {author} {\bibinfo {author} {\bibfnamefont {H.}~\bibnamefont
  {Singh}}, \bibinfo {author} {\bibfnamefont {B.~A.}\ \bibnamefont {Ware}},
  \bibinfo {author} {\bibfnamefont {R.}~\bibnamefont {Vasseur}},\ and\ \bibinfo
  {author} {\bibfnamefont {A.~J.}\ \bibnamefont {Friedman}},\ }\href
  {https://doi.org/10.1103/PhysRevLett.127.230602} {\bibfield  {journal}
  {\bibinfo  {journal} {Phys. Rev. Lett.}\ }\textbf {\bibinfo {volume} {127}},\
  \bibinfo {pages} {230602} (\bibinfo {year} {2021})}\BibitemShut {NoStop}%
\bibitem [{\citenamefont {Yang}\ \emph {et~al.}(2025)\citenamefont {Yang},
  \citenamefont {Yarloo}, \citenamefont {Zhang}, \citenamefont {M\o{}lmer},\
  and\ \citenamefont {Nielsen}}]{Yang25}%
  \BibitemOpen
  \bibfield  {author} {\bibinfo {author} {\bibfnamefont {F.}~\bibnamefont
  {Yang}}, \bibinfo {author} {\bibfnamefont {H.}~\bibnamefont {Yarloo}},
  \bibinfo {author} {\bibfnamefont {H.-C.}\ \bibnamefont {Zhang}}, \bibinfo
  {author} {\bibfnamefont {K.}~\bibnamefont {M\o{}lmer}},\ and\ \bibinfo
  {author} {\bibfnamefont {A.~E.~B.}\ \bibnamefont {Nielsen}},\ }\href
  {https://doi.org/10.1103/PhysRevB.111.144313} {\bibfield  {journal} {\bibinfo
   {journal} {Phys. Rev. B}\ }\textbf {\bibinfo {volume} {111}},\ \bibinfo
  {pages} {144313} (\bibinfo {year} {2025})}\BibitemShut {NoStop}%
\bibitem [{\citenamefont {Liu}\ \emph {et~al.}(2025)\citenamefont {Liu},
  \citenamefont {Wu}, \citenamefont {Zhang},\ and\ \citenamefont
  {Yao}}]{Liu2025}%
  \BibitemOpen
  \bibfield  {author} {\bibinfo {author} {\bibfnamefont {S.}~\bibnamefont
  {Liu}}, \bibinfo {author} {\bibfnamefont {Z.}~\bibnamefont {Wu}}, \bibinfo
  {author} {\bibfnamefont {S.-X.}\ \bibnamefont {Zhang}},\ and\ \bibinfo
  {author} {\bibfnamefont {H.}~\bibnamefont {Yao}},\ }\href
  {https://doi.org/10.1103/6722-tf9c} {\bibfield  {journal} {\bibinfo
  {journal} {Phys. Rev. B}\ }\textbf {\bibinfo {volume} {112}},\ \bibinfo
  {pages} {L020301} (\bibinfo {year} {2025})},\ \Eprint
  {https://arxiv.org/abs/2410.21386} {arXiv:2410.21386 [cond-mat.str-el]}
  \BibitemShut {NoStop}%
\bibitem [{\citenamefont {Corcoran}\ \emph {et~al.}(2025)\citenamefont
  {Corcoran}, \citenamefont {de~Leeuw},\ and\ \citenamefont
  {Pozsgay}}]{Corcoran25}%
  \BibitemOpen
  \bibfield  {author} {\bibinfo {author} {\bibfnamefont {L.}~\bibnamefont
  {Corcoran}}, \bibinfo {author} {\bibfnamefont {M.}~\bibnamefont {de~Leeuw}},\
  and\ \bibinfo {author} {\bibfnamefont {B.}~\bibnamefont {Pozsgay}},\ }\href
  {https://doi.org/10.21468/SciPostPhys.18.4.139} {\bibfield  {journal}
  {\bibinfo  {journal} {SciPost Phys.}\ }\textbf {\bibinfo {volume} {18}},\
  \bibinfo {pages} {139} (\bibinfo {year} {2025})}\BibitemShut {NoStop}%
\bibitem [{\citenamefont {Köylüoğlu}\ \emph {et~al.}(2024)\citenamefont
  {Köylüoğlu}, \citenamefont {Maskara}, \citenamefont {Feldmeier},\ and\
  \citenamefont {Lukin}}]{Naz24}%
  \BibitemOpen
  \bibfield  {author} {\bibinfo {author} {\bibfnamefont {N.~U.}\ \bibnamefont
  {Köylüoğlu}}, \bibinfo {author} {\bibfnamefont {N.}~\bibnamefont
  {Maskara}}, \bibinfo {author} {\bibfnamefont {J.}~\bibnamefont {Feldmeier}},\
  and\ \bibinfo {author} {\bibfnamefont {M.~D.}\ \bibnamefont {Lukin}},\ }\href
  {https://arxiv.org/abs/2408.02741} {\bibinfo {title} {Floquet engineering of
  interactions and entanglement in periodically driven rydberg chains}}
  (\bibinfo {year} {2024}),\ \Eprint {https://arxiv.org/abs/2408.02741}
  {arXiv:2408.02741 [quant-ph]} \BibitemShut {NoStop}%
\bibitem [{\citenamefont {Kestner}\ \emph {et~al.}(2011)\citenamefont
  {Kestner}, \citenamefont {Wang}, \citenamefont {Sau},\ and\ \citenamefont
  {Das~Sarma}}]{Kestner2011}%
  \BibitemOpen
  \bibfield  {author} {\bibinfo {author} {\bibfnamefont {J.~P.}\ \bibnamefont
  {Kestner}}, \bibinfo {author} {\bibfnamefont {B.}~\bibnamefont {Wang}},
  \bibinfo {author} {\bibfnamefont {J.~D.}\ \bibnamefont {Sau}},\ and\ \bibinfo
  {author} {\bibfnamefont {S.}~\bibnamefont {Das~Sarma}},\ }\href
  {https://doi.org/10.1103/PhysRevB.83.174409} {\bibfield  {journal} {\bibinfo
  {journal} {Phys. Rev. B}\ }\textbf {\bibinfo {volume} {83}},\ \bibinfo
  {pages} {174409} (\bibinfo {year} {2011})},\ \Eprint
  {https://arxiv.org/abs/1011.2490} {arXiv:1011.2490} \BibitemShut {NoStop}%
\bibitem [{\citenamefont {Fidkowski}\ \emph {et~al.}(2011)\citenamefont
  {Fidkowski}, \citenamefont {Lutchyn}, \citenamefont {Nayak},\ and\
  \citenamefont {Fisher}}]{Fidkowski2011}%
  \BibitemOpen
  \bibfield  {author} {\bibinfo {author} {\bibfnamefont {L.}~\bibnamefont
  {Fidkowski}}, \bibinfo {author} {\bibfnamefont {R.~M.}\ \bibnamefont
  {Lutchyn}}, \bibinfo {author} {\bibfnamefont {C.}~\bibnamefont {Nayak}},\
  and\ \bibinfo {author} {\bibfnamefont {M.~P.~A.}\ \bibnamefont {Fisher}},\
  }\href {https://doi.org/10.1103/PhysRevB.84.195436} {\bibfield  {journal}
  {\bibinfo  {journal} {Phys. Rev. B}\ }\textbf {\bibinfo {volume} {84}},\
  \bibinfo {pages} {195436} (\bibinfo {year} {2011})},\ \Eprint
  {https://arxiv.org/abs/1106.2598} {arXiv:1106.2598} \BibitemShut {NoStop}%
\bibitem [{\citenamefont {Cheng}\ and\ \citenamefont {Tu}(2011)}]{Cheng2011}%
  \BibitemOpen
  \bibfield  {author} {\bibinfo {author} {\bibfnamefont {M.}~\bibnamefont
  {Cheng}}\ and\ \bibinfo {author} {\bibfnamefont {H.-H.}\ \bibnamefont {Tu}},\
  }\href {https://doi.org/10.1103/PhysRevB.84.094503} {\bibfield  {journal}
  {\bibinfo  {journal} {Phys. Rev. B}\ }\textbf {\bibinfo {volume} {84}},\
  \bibinfo {pages} {094503} (\bibinfo {year} {2011})},\ \Eprint
  {https://arxiv.org/abs/1106.2614} {arXiv:1106.2614} \BibitemShut {NoStop}%
\bibitem [{\citenamefont {Sau}\ \emph {et~al.}(2011)\citenamefont {Sau},
  \citenamefont {Halperin}, \citenamefont {Flensberg},\ and\ \citenamefont
  {Das~Sarma}}]{Sau2011}%
  \BibitemOpen
  \bibfield  {author} {\bibinfo {author} {\bibfnamefont {J.~D.}\ \bibnamefont
  {Sau}}, \bibinfo {author} {\bibfnamefont {B.~I.}\ \bibnamefont {Halperin}},
  \bibinfo {author} {\bibfnamefont {K.}~\bibnamefont {Flensberg}},\ and\
  \bibinfo {author} {\bibfnamefont {S.}~\bibnamefont {Das~Sarma}},\ }\href
  {https://doi.org/10.1103/PhysRevB.84.144509} {\bibfield  {journal} {\bibinfo
  {journal} {Phys. Rev. B}\ }\textbf {\bibinfo {volume} {84}},\ \bibinfo
  {pages} {144509} (\bibinfo {year} {2011})},\ \Eprint
  {https://arxiv.org/abs/1106.4014} {arXiv:1106.4014} \BibitemShut {NoStop}%
\bibitem [{\citenamefont {Keselman}\ and\ \citenamefont
  {Berg}(2015)}]{Keselman2015}%
  \BibitemOpen
  \bibfield  {author} {\bibinfo {author} {\bibfnamefont {A.}~\bibnamefont
  {Keselman}}\ and\ \bibinfo {author} {\bibfnamefont {E.}~\bibnamefont
  {Berg}},\ }\href {https://doi.org/10.1103/PhysRevB.91.235309} {\bibfield
  {journal} {\bibinfo  {journal} {Phys. Rev. B}\ }\textbf {\bibinfo {volume}
  {91}},\ \bibinfo {pages} {235309} (\bibinfo {year} {2015})},\ \Eprint
  {https://arxiv.org/abs/1502.02037} {arXiv:1502.02037} \BibitemShut {NoStop}%
\bibitem [{\citenamefont {Scaffidi}\ \emph {et~al.}(2017)\citenamefont
  {Scaffidi}, \citenamefont {Parker},\ and\ \citenamefont
  {Vasseur}}]{Scaffidi2017}%
  \BibitemOpen
  \bibfield  {author} {\bibinfo {author} {\bibfnamefont {T.}~\bibnamefont
  {Scaffidi}}, \bibinfo {author} {\bibfnamefont {D.~E.}\ \bibnamefont
  {Parker}},\ and\ \bibinfo {author} {\bibfnamefont {R.}~\bibnamefont
  {Vasseur}},\ }\href {https://doi.org/10.1103/PhysRevX.7.041048} {\bibfield
  {journal} {\bibinfo  {journal} {Phys. Rev. X}\ }\textbf {\bibinfo {volume}
  {7}},\ \bibinfo {pages} {041048} (\bibinfo {year} {2017})},\ \Eprint
  {https://arxiv.org/abs/1705.01557} {arXiv:1705.01557} \BibitemShut {NoStop}%
\bibitem [{\citenamefont {Verresen}\ \emph {et~al.}(2018)\citenamefont
  {Verresen}, \citenamefont {Jones},\ and\ \citenamefont
  {Pollmann}}]{Verresen2018}%
  \BibitemOpen
  \bibfield  {author} {\bibinfo {author} {\bibfnamefont {R.}~\bibnamefont
  {Verresen}}, \bibinfo {author} {\bibfnamefont {N.~G.}\ \bibnamefont
  {Jones}},\ and\ \bibinfo {author} {\bibfnamefont {F.}~\bibnamefont
  {Pollmann}},\ }\href {https://doi.org/10.1103/PhysRevLett.120.057001}
  {\bibfield  {journal} {\bibinfo  {journal} {Phys. Rev. Lett.}\ }\textbf
  {\bibinfo {volume} {120}},\ \bibinfo {pages} {057001} (\bibinfo {year}
  {2018})},\ \Eprint {https://arxiv.org/abs/1709.03508} {arXiv:1709.03508}
  \BibitemShut {NoStop}%
\bibitem [{\citenamefont {Parker}\ \emph {et~al.}(2018)\citenamefont {Parker},
  \citenamefont {Scaffidi},\ and\ \citenamefont {Vasseur}}]{Parker2018}%
  \BibitemOpen
  \bibfield  {author} {\bibinfo {author} {\bibfnamefont {D.~E.}\ \bibnamefont
  {Parker}}, \bibinfo {author} {\bibfnamefont {T.}~\bibnamefont {Scaffidi}},\
  and\ \bibinfo {author} {\bibfnamefont {R.}~\bibnamefont {Vasseur}},\ }\href
  {https://doi.org/10.1103/PhysRevB.97.165114} {\bibfield  {journal} {\bibinfo
  {journal} {Phys. Rev. B}\ }\textbf {\bibinfo {volume} {97}},\ \bibinfo
  {pages} {165114} (\bibinfo {year} {2018})},\ \Eprint
  {https://arxiv.org/abs/1711.09106} {arXiv:1711.09106} \BibitemShut {NoStop}%
\bibitem [{\citenamefont {Verresen}\ \emph
  {et~al.}(2021{\natexlab{b}})\citenamefont {Verresen}, \citenamefont
  {Thorngren}, \citenamefont {Jones},\ and\ \citenamefont
  {Pollmann}}]{Verresen2021}%
  \BibitemOpen
  \bibfield  {author} {\bibinfo {author} {\bibfnamefont {R.}~\bibnamefont
  {Verresen}}, \bibinfo {author} {\bibfnamefont {R.}~\bibnamefont {Thorngren}},
  \bibinfo {author} {\bibfnamefont {N.~G.}\ \bibnamefont {Jones}},\ and\
  \bibinfo {author} {\bibfnamefont {F.}~\bibnamefont {Pollmann}},\ }\bibfield
  {journal} {\bibinfo  {journal} {Physical Review X}\ }\textbf {\bibinfo
  {volume} {11}},\ \href {https://doi.org/10.1103/physrevx.11.041059}
  {10.1103/physrevx.11.041059} (\bibinfo {year}
  {2021}{\natexlab{b}})\BibitemShut {NoStop}%
\bibitem [{\citenamefont {Thorngren}\ \emph {et~al.}(2021)\citenamefont
  {Thorngren}, \citenamefont {Vishwanath},\ and\ \citenamefont
  {Verresen}}]{Thorngren2021}%
  \BibitemOpen
  \bibfield  {author} {\bibinfo {author} {\bibfnamefont {R.}~\bibnamefont
  {Thorngren}}, \bibinfo {author} {\bibfnamefont {A.}~\bibnamefont
  {Vishwanath}},\ and\ \bibinfo {author} {\bibfnamefont {R.}~\bibnamefont
  {Verresen}},\ }\href {https://doi.org/10.1103/PhysRevB.104.075132} {\bibfield
   {journal} {\bibinfo  {journal} {Phys. Rev. B}\ }\textbf {\bibinfo {volume}
  {104}},\ \bibinfo {pages} {075132} (\bibinfo {year} {2021})},\ \Eprint
  {https://arxiv.org/abs/2008.06638} {arXiv:2008.06638} \BibitemShut {NoStop}%
\bibitem [{\citenamefont {Li}\ \emph {et~al.}(2024{\natexlab{a}})\citenamefont
  {Li}, \citenamefont {Oshikawa},\ and\ \citenamefont {Zheng}}]{Li2024SPT}%
  \BibitemOpen
  \bibfield  {author} {\bibinfo {author} {\bibfnamefont {L.}~\bibnamefont
  {Li}}, \bibinfo {author} {\bibfnamefont {M.}~\bibnamefont {Oshikawa}},\ and\
  \bibinfo {author} {\bibfnamefont {Y.}~\bibnamefont {Zheng}},\ }\href
  {https://doi.org/10.21468/SciPostPhys.17.1.013} {\bibfield  {journal}
  {\bibinfo  {journal} {SciPost Phys.}\ }\textbf {\bibinfo {volume} {17}},\
  \bibinfo {pages} {013} (\bibinfo {year} {2024}{\natexlab{a}})},\ \Eprint
  {https://arxiv.org/abs/2204.03131} {arXiv:2204.03131} \BibitemShut {NoStop}%
\bibitem [{\citenamefont {Wen}\ and\ \citenamefont {Potter}(2025)}]{Wen2025}%
  \BibitemOpen
  \bibfield  {author} {\bibinfo {author} {\bibfnamefont {R.}~\bibnamefont
  {Wen}}\ and\ \bibinfo {author} {\bibfnamefont {A.~C.}\ \bibnamefont
  {Potter}},\ }\href {https://doi.org/10.1103/PhysRevB.111.115161} {\bibfield
  {journal} {\bibinfo  {journal} {Phys. Rev. B}\ }\textbf {\bibinfo {volume}
  {111}},\ \bibinfo {pages} {115161} (\bibinfo {year} {2025})},\ \Eprint
  {https://arxiv.org/abs/2311.00050} {arXiv:2311.00050} \BibitemShut {NoStop}%
\bibitem [{\citenamefont {Alcaraz}\ and\ \citenamefont
  {Bariev}(1999)}]{Alcaraz99}%
  \BibitemOpen
  \bibfield  {author} {\bibinfo {author} {\bibfnamefont {F.~C.}\ \bibnamefont
  {Alcaraz}}\ and\ \bibinfo {author} {\bibfnamefont {R.~Z.}\ \bibnamefont
  {Bariev}},\ }\href {https://arxiv.org/abs/cond-mat/9904042} {\bibinfo {title}
  {An exactly solvable constrained xxz chain}} (\bibinfo {year} {1999}),\
  \Eprint {https://arxiv.org/abs/cond-mat/9904042} {arXiv:cond-mat/9904042
  [cond-mat.stat-mech]} \BibitemShut {NoStop}%
\bibitem [{\citenamefont {Wang}\ \emph {et~al.}(2025)\citenamefont {Wang},
  \citenamefont {Li},\ and\ \citenamefont {Li}}]{Wang2025}%
  \BibitemOpen
  \bibfield  {author} {\bibinfo {author} {\bibfnamefont {H.}~\bibnamefont
  {Wang}}, \bibinfo {author} {\bibfnamefont {X.}~\bibnamefont {Li}},\ and\
  \bibinfo {author} {\bibfnamefont {C.}~\bibnamefont {Li}},\ }\href
  {https://doi.org/10.1038/s41467-025-65652-9} {\bibfield  {journal} {\bibinfo
  {journal} {Nat. Commun.}\ }\textbf {\bibinfo {volume} {16}},\ \bibinfo
  {pages} {10584} (\bibinfo {year} {2025})},\ \Eprint
  {https://arxiv.org/abs/2505.12979} {arXiv:2505.12979} \BibitemShut {NoStop}%
\bibitem [{\citenamefont {Baier}\ \emph {et~al.}(2016)\citenamefont {Baier},
  \citenamefont {Mark}, \citenamefont {Petter}, \citenamefont {Aikawa},
  \citenamefont {Chomaz}, \citenamefont {Cai}, \citenamefont {Baranov},
  \citenamefont {Zoller},\ and\ \citenamefont {Ferlaino}}]{Baier2016}%
  \BibitemOpen
  \bibfield  {author} {\bibinfo {author} {\bibfnamefont {S.}~\bibnamefont
  {Baier}}, \bibinfo {author} {\bibfnamefont {M.~J.}\ \bibnamefont {Mark}},
  \bibinfo {author} {\bibfnamefont {D.}~\bibnamefont {Petter}}, \bibinfo
  {author} {\bibfnamefont {K.}~\bibnamefont {Aikawa}}, \bibinfo {author}
  {\bibfnamefont {L.}~\bibnamefont {Chomaz}}, \bibinfo {author} {\bibfnamefont
  {Z.}~\bibnamefont {Cai}}, \bibinfo {author} {\bibfnamefont {M.}~\bibnamefont
  {Baranov}}, \bibinfo {author} {\bibfnamefont {P.}~\bibnamefont {Zoller}},\
  and\ \bibinfo {author} {\bibfnamefont {F.}~\bibnamefont {Ferlaino}},\ }\href
  {https://doi.org/10.1126/science.aac9812} {\bibfield  {journal} {\bibinfo
  {journal} {Science}\ }\textbf {\bibinfo {volume} {352}},\ \bibinfo {pages}
  {201–205} (\bibinfo {year} {2016})}\BibitemShut {NoStop}%
\bibitem [{\citenamefont {Su}\ \emph {et~al.}(2023)\citenamefont {Su},
  \citenamefont {Douglas}, \citenamefont {Szurek}, \citenamefont {Groth},
  \citenamefont {Ozturk}, \citenamefont {Krahn}, \citenamefont {Hébert},
  \citenamefont {Phelps}, \citenamefont {Ebadi}, \citenamefont {Dickerson},
  \citenamefont {Ferlaino}, \citenamefont {Marković},\ and\ \citenamefont
  {Greiner}}]{Su2023}%
  \BibitemOpen
  \bibfield  {author} {\bibinfo {author} {\bibfnamefont {L.}~\bibnamefont
  {Su}}, \bibinfo {author} {\bibfnamefont {A.}~\bibnamefont {Douglas}},
  \bibinfo {author} {\bibfnamefont {M.}~\bibnamefont {Szurek}}, \bibinfo
  {author} {\bibfnamefont {R.}~\bibnamefont {Groth}}, \bibinfo {author}
  {\bibfnamefont {S.~F.}\ \bibnamefont {Ozturk}}, \bibinfo {author}
  {\bibfnamefont {A.}~\bibnamefont {Krahn}}, \bibinfo {author} {\bibfnamefont
  {A.~H.}\ \bibnamefont {Hébert}}, \bibinfo {author} {\bibfnamefont {G.~A.}\
  \bibnamefont {Phelps}}, \bibinfo {author} {\bibfnamefont {S.}~\bibnamefont
  {Ebadi}}, \bibinfo {author} {\bibfnamefont {S.}~\bibnamefont {Dickerson}},
  \bibinfo {author} {\bibfnamefont {F.}~\bibnamefont {Ferlaino}}, \bibinfo
  {author} {\bibfnamefont {O.}~\bibnamefont {Marković}},\ and\ \bibinfo
  {author} {\bibfnamefont {M.}~\bibnamefont {Greiner}},\ }\href
  {https://doi.org/10.1038/s41586-023-06614-3} {\bibfield  {journal} {\bibinfo
  {journal} {Nature}\ }\textbf {\bibinfo {volume} {622}},\ \bibinfo {pages}
  {724–729} (\bibinfo {year} {2023})}\BibitemShut {NoStop}%
\bibitem [{\citenamefont {Su}\ \emph {et~al.}(2026)\citenamefont {Su},
  \citenamefont {Sahay}, \citenamefont {Szurek}, \citenamefont {Douglas},
  \citenamefont {Marković}, \citenamefont {Dag}, \citenamefont {Verresen},\
  and\ \citenamefont {Greiner}}]{Su2026}%
  \BibitemOpen
  \bibfield  {author} {\bibinfo {author} {\bibfnamefont {L.}~\bibnamefont
  {Su}}, \bibinfo {author} {\bibfnamefont {R.}~\bibnamefont {Sahay}}, \bibinfo
  {author} {\bibfnamefont {M.}~\bibnamefont {Szurek}}, \bibinfo {author}
  {\bibfnamefont {A.}~\bibnamefont {Douglas}}, \bibinfo {author} {\bibfnamefont
  {O.}~\bibnamefont {Marković}}, \bibinfo {author} {\bibfnamefont {C.~B.}\
  \bibnamefont {Dag}}, \bibinfo {author} {\bibfnamefont {R.}~\bibnamefont
  {Verresen}},\ and\ \bibinfo {author} {\bibfnamefont {M.}~\bibnamefont
  {Greiner}},\ }\href {https://doi.org/10.1038/s41567-026-03381-6} {\bibfield
  {journal} {\bibinfo  {journal} {Nature Physics}\ }\textbf {\bibinfo {volume}
  {22}},\ \bibinfo {pages} {1672–1678} (\bibinfo {year} {2026})}\BibitemShut
  {NoStop}%
\bibitem [{\citenamefont {Ogata}\ and\ \citenamefont
  {Shiba}(1990)}]{ogata1990}%
  \BibitemOpen
  \bibfield  {author} {\bibinfo {author} {\bibfnamefont {M.}~\bibnamefont
  {Ogata}}\ and\ \bibinfo {author} {\bibfnamefont {H.}~\bibnamefont {Shiba}},\
  }\href {https://doi.org/10.1103/PhysRevB.41.2326} {\bibfield  {journal}
  {\bibinfo  {journal} {Phys. Rev. B}\ }\textbf {\bibinfo {volume} {41}},\
  \bibinfo {pages} {2326} (\bibinfo {year} {1990})}\BibitemShut {NoStop}%
\bibitem [{\citenamefont {Zhang}\ \emph {et~al.}(1997)\citenamefont {Zhang},
  \citenamefont {Karbach}, \citenamefont {M{\"u}ller},\ and\ \citenamefont
  {Stolze}}]{zhang1997}%
  \BibitemOpen
  \bibfield  {author} {\bibinfo {author} {\bibfnamefont {S.}~\bibnamefont
  {Zhang}}, \bibinfo {author} {\bibfnamefont {M.}~\bibnamefont {Karbach}},
  \bibinfo {author} {\bibfnamefont {G.}~\bibnamefont {M{\"u}ller}},\ and\
  \bibinfo {author} {\bibfnamefont {J.}~\bibnamefont {Stolze}},\ }\href
  {https://doi.org/10.1103/PhysRevB.55.6491} {\bibfield  {journal} {\bibinfo
  {journal} {Phys. Rev. B}\ }\textbf {\bibinfo {volume} {55}},\ \bibinfo
  {pages} {6491} (\bibinfo {year} {1997})}\BibitemShut {NoStop}%
\bibitem [{\citenamefont {Kruis}\ \emph {et~al.}(2004)\citenamefont {Kruis},
  \citenamefont {McCulloch}, \citenamefont {Nussinov},\ and\ \citenamefont
  {Zaanen}}]{kruis2004}%
  \BibitemOpen
  \bibfield  {author} {\bibinfo {author} {\bibfnamefont {H.~V.}\ \bibnamefont
  {Kruis}}, \bibinfo {author} {\bibfnamefont {I.~P.}\ \bibnamefont
  {McCulloch}}, \bibinfo {author} {\bibfnamefont {Z.}~\bibnamefont
  {Nussinov}},\ and\ \bibinfo {author} {\bibfnamefont {J.}~\bibnamefont
  {Zaanen}},\ }\href {https://doi.org/10.1103/PhysRevB.70.075109} {\bibfield
  {journal} {\bibinfo  {journal} {Phys. Rev. B}\ }\textbf {\bibinfo {volume}
  {70}},\ \bibinfo {pages} {075109} (\bibinfo {year} {2004})}\BibitemShut
  {NoStop}%
\bibitem [{\citenamefont {Pollmann}\ \emph {et~al.}(2012)\citenamefont
  {Pollmann}, \citenamefont {Berg}, \citenamefont {Turner},\ and\ \citenamefont
  {Oshikawa}}]{Pollmann2012}%
  \BibitemOpen
  \bibfield  {author} {\bibinfo {author} {\bibfnamefont {F.}~\bibnamefont
  {Pollmann}}, \bibinfo {author} {\bibfnamefont {E.}~\bibnamefont {Berg}},
  \bibinfo {author} {\bibfnamefont {A.~M.}\ \bibnamefont {Turner}},\ and\
  \bibinfo {author} {\bibfnamefont {M.}~\bibnamefont {Oshikawa}},\ }\bibfield
  {journal} {\bibinfo  {journal} {Physical Review B}\ }\textbf {\bibinfo
  {volume} {85}},\ \href {https://doi.org/10.1103/physrevb.85.075125}
  {10.1103/physrevb.85.075125} (\bibinfo {year} {2012})\BibitemShut {NoStop}%
\bibitem [{\citenamefont {Schuch}\ \emph {et~al.}(2011)\citenamefont {Schuch},
  \citenamefont {Pérez-García},\ and\ \citenamefont {Cirac}}]{Schuch2011}%
  \BibitemOpen
  \bibfield  {author} {\bibinfo {author} {\bibfnamefont {N.}~\bibnamefont
  {Schuch}}, \bibinfo {author} {\bibfnamefont {D.}~\bibnamefont
  {Pérez-García}},\ and\ \bibinfo {author} {\bibfnamefont {I.}~\bibnamefont
  {Cirac}},\ }\bibfield  {journal} {\bibinfo  {journal} {Physical Review B}\
  }\textbf {\bibinfo {volume} {84}},\ \href
  {https://doi.org/10.1103/physrevb.84.165139} {10.1103/physrevb.84.165139}
  (\bibinfo {year} {2011})\BibitemShut {NoStop}%
\bibitem [{\citenamefont {Chen}\ \emph {et~al.}(2011)\citenamefont {Chen},
  \citenamefont {Gu},\ and\ \citenamefont {Wen}}]{Chen2011}%
  \BibitemOpen
  \bibfield  {author} {\bibinfo {author} {\bibfnamefont {X.}~\bibnamefont
  {Chen}}, \bibinfo {author} {\bibfnamefont {Z.-C.}\ \bibnamefont {Gu}},\ and\
  \bibinfo {author} {\bibfnamefont {X.-G.}\ \bibnamefont {Wen}},\ }\bibfield
  {journal} {\bibinfo  {journal} {Physical Review B}\ }\textbf {\bibinfo
  {volume} {84}},\ \href {https://doi.org/10.1103/physrevb.84.235128}
  {10.1103/physrevb.84.235128} (\bibinfo {year} {2011})\BibitemShut {NoStop}%
\bibitem [{Note1()}]{Note1}%
  \BibitemOpen
  \bibinfo {note} {We note that the statistics does not affect the physics at
  hand, so $a_{j,s}$ can be taken to be a hardcore boson or
  fermion.}\BibitemShut {Stop}%
\bibitem [{sup()}]{supp}%
  \BibitemOpen
  \href@noop {} {}\bibinfo {note} {See Supplemental Material for details on the
  exactly solvable line, the dilute hole limit, and the bosonized analysis of
  the three-leg model.}\BibitemShut {Stop}%
\bibitem [{\citenamefont {White}(1992)}]{White1992}%
  \BibitemOpen
  \bibfield  {author} {\bibinfo {author} {\bibfnamefont {S.~R.}\ \bibnamefont
  {White}},\ }\href {https://doi.org/10.1103/PhysRevLett.69.2863} {\bibfield
  {journal} {\bibinfo  {journal} {Phys. Rev. Lett.}\ }\textbf {\bibinfo
  {volume} {69}},\ \bibinfo {pages} {2863} (\bibinfo {year}
  {1992})}\BibitemShut {NoStop}%
\bibitem [{\citenamefont {White}(1993)}]{White1993}%
  \BibitemOpen
  \bibfield  {author} {\bibinfo {author} {\bibfnamefont {S.~R.}\ \bibnamefont
  {White}},\ }\href {https://doi.org/10.1103/PhysRevB.48.10345} {\bibfield
  {journal} {\bibinfo  {journal} {Phys. Rev. B}\ }\textbf {\bibinfo {volume}
  {48}},\ \bibinfo {pages} {10345} (\bibinfo {year} {1993})}\BibitemShut
  {NoStop}%
\bibitem [{\citenamefont {Hauschild}\ and\ \citenamefont
  {Pollmann}(2018)}]{Hauschild2018}%
  \BibitemOpen
  \bibfield  {author} {\bibinfo {author} {\bibfnamefont {J.}~\bibnamefont
  {Hauschild}}\ and\ \bibinfo {author} {\bibfnamefont {F.}~\bibnamefont
  {Pollmann}},\ }\href {https://doi.org/10.21468/SciPostPhysLectNotes.5}
  {\bibfield  {journal} {\bibinfo  {journal} {SciPost Phys. Lect. Notes}\ ,\
  \bibinfo {pages} {5}} (\bibinfo {year} {2018})}\BibitemShut {NoStop}%
\bibitem [{\citenamefont {Hauschild}\ \emph {et~al.}(2024)\citenamefont
  {Hauschild}, \citenamefont {Unfried}, \citenamefont {Anand}, \citenamefont
  {Andrews}, \citenamefont {Bintz}, \citenamefont {Borla}, \citenamefont
  {Divic}, \citenamefont {Drescher}, \citenamefont {Geiger}, \citenamefont
  {Hefel}, \citenamefont {Hémery}, \citenamefont {Kadow}, \citenamefont
  {Kemp}, \citenamefont {Kirchner}, \citenamefont {Liu}, \citenamefont
  {Möller}, \citenamefont {Parker}, \citenamefont {Rader}, \citenamefont
  {Romen}, \citenamefont {Scalet}, \citenamefont {Schoonderwoerd},
  \citenamefont {Schulz}, \citenamefont {Soejima}, \citenamefont {Thoma},
  \citenamefont {Wu}, \citenamefont {Zechmann}, \citenamefont {Zweng},
  \citenamefont {Mong}, \citenamefont {Zaletel},\ and\ \citenamefont
  {Pollmann}}]{tenpy2024}%
  \BibitemOpen
  \bibfield  {author} {\bibinfo {author} {\bibfnamefont {J.}~\bibnamefont
  {Hauschild}}, \bibinfo {author} {\bibfnamefont {J.}~\bibnamefont {Unfried}},
  \bibinfo {author} {\bibfnamefont {S.}~\bibnamefont {Anand}}, \bibinfo
  {author} {\bibfnamefont {B.}~\bibnamefont {Andrews}}, \bibinfo {author}
  {\bibfnamefont {M.}~\bibnamefont {Bintz}}, \bibinfo {author} {\bibfnamefont
  {U.}~\bibnamefont {Borla}}, \bibinfo {author} {\bibfnamefont
  {S.}~\bibnamefont {Divic}}, \bibinfo {author} {\bibfnamefont
  {M.}~\bibnamefont {Drescher}}, \bibinfo {author} {\bibfnamefont
  {J.}~\bibnamefont {Geiger}}, \bibinfo {author} {\bibfnamefont
  {M.}~\bibnamefont {Hefel}}, \bibinfo {author} {\bibfnamefont
  {K.}~\bibnamefont {Hémery}}, \bibinfo {author} {\bibfnamefont
  {W.}~\bibnamefont {Kadow}}, \bibinfo {author} {\bibfnamefont
  {J.}~\bibnamefont {Kemp}}, \bibinfo {author} {\bibfnamefont {N.}~\bibnamefont
  {Kirchner}}, \bibinfo {author} {\bibfnamefont {V.~S.}\ \bibnamefont {Liu}},
  \bibinfo {author} {\bibfnamefont {G.}~\bibnamefont {Möller}}, \bibinfo
  {author} {\bibfnamefont {D.}~\bibnamefont {Parker}}, \bibinfo {author}
  {\bibfnamefont {M.}~\bibnamefont {Rader}}, \bibinfo {author} {\bibfnamefont
  {A.}~\bibnamefont {Romen}}, \bibinfo {author} {\bibfnamefont
  {S.}~\bibnamefont {Scalet}}, \bibinfo {author} {\bibfnamefont
  {L.}~\bibnamefont {Schoonderwoerd}}, \bibinfo {author} {\bibfnamefont
  {M.}~\bibnamefont {Schulz}}, \bibinfo {author} {\bibfnamefont
  {T.}~\bibnamefont {Soejima}}, \bibinfo {author} {\bibfnamefont
  {P.}~\bibnamefont {Thoma}}, \bibinfo {author} {\bibfnamefont
  {Y.}~\bibnamefont {Wu}}, \bibinfo {author} {\bibfnamefont {P.}~\bibnamefont
  {Zechmann}}, \bibinfo {author} {\bibfnamefont {L.}~\bibnamefont {Zweng}},
  \bibinfo {author} {\bibfnamefont {R.~S.~K.}\ \bibnamefont {Mong}}, \bibinfo
  {author} {\bibfnamefont {M.~P.}\ \bibnamefont {Zaletel}},\ and\ \bibinfo
  {author} {\bibfnamefont {F.}~\bibnamefont {Pollmann}},\ }\href
  {https://doi.org/10.21468/SciPostPhysCodeb.41} {\bibfield  {journal}
  {\bibinfo  {journal} {SciPost Phys. Codebases}\ ,\ \bibinfo {pages} {41}}
  (\bibinfo {year} {2024})}\BibitemShut {NoStop}%
\bibitem [{\citenamefont {Pollmann}\ \emph {et~al.}(2009)\citenamefont
  {Pollmann}, \citenamefont {Mukerjee}, \citenamefont {Turner},\ and\
  \citenamefont {Moore}}]{Pollmann2009}%
  \BibitemOpen
  \bibfield  {author} {\bibinfo {author} {\bibfnamefont {F.}~\bibnamefont
  {Pollmann}}, \bibinfo {author} {\bibfnamefont {S.}~\bibnamefont {Mukerjee}},
  \bibinfo {author} {\bibfnamefont {A.~M.}\ \bibnamefont {Turner}},\ and\
  \bibinfo {author} {\bibfnamefont {J.~E.}\ \bibnamefont {Moore}},\ }\href
  {https://doi.org/10.1103/PhysRevLett.102.255701} {\bibfield  {journal}
  {\bibinfo  {journal} {Phys. Rev. Lett.}\ }\textbf {\bibinfo {volume} {102}},\
  \bibinfo {pages} {255701} (\bibinfo {year} {2009})}\BibitemShut {NoStop}%
\bibitem [{\citenamefont {Bauer}\ \emph {et~al.}(2013)\citenamefont {Bauer},
  \citenamefont {Huijse}, \citenamefont {Berg}, \citenamefont {Troyer},\ and\
  \citenamefont {Schoutens}}]{Bauer2013}%
  \BibitemOpen
  \bibfield  {author} {\bibinfo {author} {\bibfnamefont {B.}~\bibnamefont
  {Bauer}}, \bibinfo {author} {\bibfnamefont {L.}~\bibnamefont {Huijse}},
  \bibinfo {author} {\bibfnamefont {E.}~\bibnamefont {Berg}}, \bibinfo {author}
  {\bibfnamefont {M.}~\bibnamefont {Troyer}},\ and\ \bibinfo {author}
  {\bibfnamefont {K.}~\bibnamefont {Schoutens}},\ }\bibfield  {journal}
  {\bibinfo  {journal} {Physical Review B}\ }\textbf {\bibinfo {volume} {87}},\
  \href {https://doi.org/10.1103/physrevb.87.165145}
  {10.1103/physrevb.87.165145} (\bibinfo {year} {2013})\BibitemShut {NoStop}%
\bibitem [{\citenamefont {Li}\ \emph {et~al.}(2024{\natexlab{b}})\citenamefont
  {Li}, \citenamefont {Liu}, \citenamefont {Wang}, \citenamefont {Zhang},
  \citenamefont {Li}, \citenamefont {Zhai},\ and\ \citenamefont {Gu}}]{Li2024}%
  \BibitemOpen
  \bibfield  {author} {\bibinfo {author} {\bibfnamefont {C.}~\bibnamefont
  {Li}}, \bibinfo {author} {\bibfnamefont {S.}~\bibnamefont {Liu}}, \bibinfo
  {author} {\bibfnamefont {H.}~\bibnamefont {Wang}}, \bibinfo {author}
  {\bibfnamefont {W.}~\bibnamefont {Zhang}}, \bibinfo {author} {\bibfnamefont
  {Z.-X.}\ \bibnamefont {Li}}, \bibinfo {author} {\bibfnamefont
  {H.}~\bibnamefont {Zhai}},\ and\ \bibinfo {author} {\bibfnamefont
  {Y.}~\bibnamefont {Gu}},\ }\href
  {https://doi.org/10.1103/PhysRevLett.133.223401} {\bibfield  {journal}
  {\bibinfo  {journal} {Phys. Rev. Lett.}\ }\textbf {\bibinfo {volume} {133}},\
  \bibinfo {pages} {223401} (\bibinfo {year} {2024}{\natexlab{b}})},\ \Eprint
  {https://arxiv.org/abs/2407.08194} {arXiv:2407.08194} \BibitemShut {NoStop}%
\bibitem [{\citenamefont {Wang}\ \emph {et~al.}(2026)\citenamefont {Wang},
  \citenamefont {Li}, \citenamefont {Liu}, \citenamefont {Gu},\ and\
  \citenamefont {Li}}]{Wang2026}%
  \BibitemOpen
  \bibfield  {author} {\bibinfo {author} {\bibfnamefont {H.}~\bibnamefont
  {Wang}}, \bibinfo {author} {\bibfnamefont {X.}~\bibnamefont {Li}}, \bibinfo
  {author} {\bibfnamefont {S.}~\bibnamefont {Liu}}, \bibinfo {author}
  {\bibfnamefont {Y.}~\bibnamefont {Gu}},\ and\ \bibinfo {author}
  {\bibfnamefont {C.}~\bibnamefont {Li}},\ }\href@noop {} {\bibinfo {title}
  {Unleashing emergent fermions with {R}ydberg atom simulators}} (\bibinfo
  {year} {2026}),\ \Eprint {https://arxiv.org/abs/2606.19444} {arXiv:2606.19444
  [cond-mat.quant-gas]} \BibitemShut {NoStop}%
\bibitem [{\citenamefont {Zaletel}\ \emph {et~al.}(2015)\citenamefont
  {Zaletel}, \citenamefont {Mong}, \citenamefont {Karrasch}, \citenamefont
  {Moore},\ and\ \citenamefont {Pollmann}}]{Zaletel2015}%
  \BibitemOpen
  \bibfield  {author} {\bibinfo {author} {\bibfnamefont {M.~P.}\ \bibnamefont
  {Zaletel}}, \bibinfo {author} {\bibfnamefont {R.~S.~K.}\ \bibnamefont
  {Mong}}, \bibinfo {author} {\bibfnamefont {C.}~\bibnamefont {Karrasch}},
  \bibinfo {author} {\bibfnamefont {J.~E.}\ \bibnamefont {Moore}},\ and\
  \bibinfo {author} {\bibfnamefont {F.}~\bibnamefont {Pollmann}},\ }\href
  {https://doi.org/10.1103/PhysRevB.91.165112} {\bibfield  {journal} {\bibinfo
  {journal} {Phys. Rev. B}\ }\textbf {\bibinfo {volume} {91}},\ \bibinfo
  {pages} {165112} (\bibinfo {year} {2015})}\BibitemShut {NoStop}%
\bibitem [{\citenamefont {Spohn}(1991)}]{Spohn1991}%
  \BibitemOpen
  \bibfield  {author} {\bibinfo {author} {\bibfnamefont {H.}~\bibnamefont
  {Spohn}},\ }\href {https://doi.org/10.1007/978-3-642-84371-6} {\emph
  {\bibinfo {title} {Large Scale Dynamics of Interacting Particles}}}\
  (\bibinfo  {publisher} {Springer},\ \bibinfo {address} {Berlin, Heidelberg},\
  \bibinfo {year} {1991})\BibitemShut {NoStop}%
\bibitem [{\citenamefont {Hilker}\ \emph {et~al.}(2017)\citenamefont {Hilker},
  \citenamefont {Salomon}, \citenamefont {Grusdt}, \citenamefont {Omran},
  \citenamefont {Boll}, \citenamefont {Demler}, \citenamefont {Bloch},\ and\
  \citenamefont {Gross}}]{Hilker2017}%
  \BibitemOpen
  \bibfield  {author} {\bibinfo {author} {\bibfnamefont {T.~A.}\ \bibnamefont
  {Hilker}}, \bibinfo {author} {\bibfnamefont {G.}~\bibnamefont {Salomon}},
  \bibinfo {author} {\bibfnamefont {F.}~\bibnamefont {Grusdt}}, \bibinfo
  {author} {\bibfnamefont {A.}~\bibnamefont {Omran}}, \bibinfo {author}
  {\bibfnamefont {M.}~\bibnamefont {Boll}}, \bibinfo {author} {\bibfnamefont
  {E.}~\bibnamefont {Demler}}, \bibinfo {author} {\bibfnamefont
  {I.}~\bibnamefont {Bloch}},\ and\ \bibinfo {author} {\bibfnamefont
  {C.}~\bibnamefont {Gross}},\ }\href {https://doi.org/10.1126/science.aam8990}
  {\bibfield  {journal} {\bibinfo  {journal} {Science}\ }\textbf {\bibinfo
  {volume} {357}},\ \bibinfo {pages} {484–487} (\bibinfo {year}
  {2017})}\BibitemShut {NoStop}%
\bibitem [{\citenamefont {Li}\ \emph {et~al.}(2025)\citenamefont {Li},
  \citenamefont {Oshikawa},\ and\ \citenamefont {Zheng}}]{Li2025}%
  \BibitemOpen
  \bibfield  {author} {\bibinfo {author} {\bibfnamefont {L.}~\bibnamefont
  {Li}}, \bibinfo {author} {\bibfnamefont {M.}~\bibnamefont {Oshikawa}},\ and\
  \bibinfo {author} {\bibfnamefont {Y.}~\bibnamefont {Zheng}},\ }\bibfield
  {journal} {\bibinfo  {journal} {SciPost Physics}\ }\textbf {\bibinfo {volume}
  {18}},\ \href {https://doi.org/10.21468/scipostphys.18.5.153}
  {10.21468/scipostphys.18.5.153} (\bibinfo {year} {2025})\BibitemShut
  {NoStop}%
\bibitem [{\citenamefont {Giamarchi}(2004)}]{Giamarchi2004}%
  \BibitemOpen
  \bibfield  {author} {\bibinfo {author} {\bibfnamefont {T.}~\bibnamefont
  {Giamarchi}},\ }\href
  {https://doi.org/10.1093/acprof:oso/9780198525004.001.0001} {\emph {\bibinfo
  {title} {Quantum Physics in One Dimension}}}\ (\bibinfo  {publisher} {Oxford
  University Press},\ \bibinfo {address} {Oxford},\ \bibinfo {year}
  {2004})\BibitemShut {NoStop}%
\end{thebibliography}%

\appendix

\section{End Matter}

\textit{Perturbation theory---}%
We here demonstrate that Eq.~\eqref{eq:PXP_Eff} arises at second order in the perturbative expansion of Eq.~\eqref{eq:PXP} and provide the explicit form of $H_{\mathrm{PXYP}}'$ in the two-leg geometry.
Consider a Rydberg array of qubits on an arbitrary lattice and let $G=(V,E)$ be the \emph{blockade graph}, where the vertices, $V$, are lattice sites and the edges, $E$, connect sites which are forbidden to be simultaneously occupied due to the Rydberg blockade. Explicitly, for a site $i$, defining $n_i = \ket{1}_i\bra{1}_i$, $P_i = \prod_{\{ i, j \} \in E} (1-n_j)$, and $T_+ = \sum_i \sigma_i^+ P_i$, the PXP Hamiltonian is given by,
\begin{align}
    H_{\mathrm{PXP}} = -\frac{\Omega}{2}(T_+ + T_+^\dagger) - \delta N,
\end{align}
where $\delta$ is the detuning. Now, in the strong detuning limit, $\delta / \Omega \gg 1$, one readily confirms that the effective Hamiltonian up to second order in $\Omega/\delta$ is given by,
\begin{align}
    H_{\mathrm{PXYP}}' = -\delta N - t [T_+ , T_+^\dagger] \label{eq:app_H2}
\end{align}
where $t = \Omega^2 / 4\delta$ and $N = \sum_i n_i$. Expanding out the commutator requires the evaluation of $[\sigma_i^+ P_i , P_j \sigma^-_j]$; this is only non-zero if $i=j$ or $\{i,j\} \in E$. One then finds,
\begin{subequations}
\begin{align}
    \tag{\theparentequation} [T_+, T_+^\dagger] &= 2N - V + K \\
    V &= \sum_i P_i  \\
    K &= P \sum_{\{i,j\} \in E} (\sigma_i^+ \sigma_j^- + h.c. )  P \, ,
\end{align}
\end{subequations}
where $P$ is the projector onto the Rydberg constraint satisfying subspace.
Here, $2N-V$ ($K$) arises from commutators of terms on the same (nearest-neighbor) sites. Altogether,
\begin{align}
    H_{\mathrm{PXYP}}' = H_{\mathrm{PXYP}} -(\delta +2t) N + tV  \, ,
\end{align}
where we used the fact that $H_{\mathrm{PXYP}}=-tK$ is exactly the PXYP Hamiltonian.

As we did for $H_{\mathrm{PXYP}}$ in Eq.~\eqref{eq:hamiltonian-base}, we can consider a generalized version of $H_{\mathrm{PXYP}}'$ by allowing for independent couplings for the hopping, flavor flips, and density-density interaction. 
Using the pseudospin notation of Eq.~\eqref{eq:hamiltonian-base}, and working in a fixed $N$ sector with periodic boundary conditions, 
we define
\begin{align}
    \begin{split}
    H(t_\parallel, t_\perp,\lambda) = &H(t_\parallel, t_\perp) + 2\lambda\sum_i n_i n_{i+1} \\
    &+ \lambda\sum_{i,s} (1-n_i) n_{i-1,s}n_{i+1,s} \, ,
    \end{split}
\end{align}
where we have dropped an overall constant and used the Rydberg constraint $n_{i,s}n_{i+1,s} = 0$ in rewriting $V$. In the main text and in Fig.~\ref{fig:igSPT}, we discuss $H(t_\parallel, t_\perp) \equiv H(t_\parallel, t_\perp,\lambda = 0)$ and $H'(t_\parallel, t_\perp) \equiv H(t_\parallel, t_\perp,\lambda = 1)$.

\textit{Transition line in the dilute particle limit---}%
Let us consider the PXYP model in the limit of small filling and attempt to address the position and nature of the transition from the topological Luttinger liquid phase to the trivial phase. We again make use of the fact that the line $t_\perp = 0$ is exactly solvable and maps to a problem of nearest-neighbor hopping hardcore bosons \cite{supp}, which is Jordan-Wigner dual to free fermions. Using the squeezed-space representation, let us label the particle positions as $x_1 < x_2 < \dots < x_{N_p}$ and their corresponding flavors as $\sigma^z_a = \pm 1$. We define $d_a = (1+\sigma^z_a \sigma^z_{a+1})/2$, such that $d_a = +1$ indicates an \emph{improper} bond, and define the total number of improper bonds as $M = \sum_{a=1}^{N_p-1} d_a$. A state with exact AFM order in squeezed space has $M=0$. Now, an improper bond $d_a=1$ effectively traps a hole in the physical, unsqueezed space (as otherwise the Rydberg blockade would be violated). As shown in \cite{supp}, the ground state energy in a fixed $M$ sector for $t_\perp = 0$ is given by,
\begin{align}
	E(M) = -\frac{2t_\parallel (L-M)}{\pi} \sin\left( \frac{\pi N_p}{L-M} \right)
\end{align}
in the thermodynamic limit.

For $t_\perp = 0$, the ground state has $M=0$. As we turn on $t_\perp$ and approach the transition to the trivial phase, a finite density of improper bonds will be excited in the ground state. Let $\nu = N_p/L$ denote the filling and let the ratio of improper bonds to particles, $f=M/N_p$, be an $O(1)$ quantity, such that $f\nu$ is a small number as we take $\nu \to 0$. In this limit, we have for the energy density,
\begin{align}
	E(M)/L = -2\tpar \nu + \pi^2 \tpar \nu^3/3 + 2\pi^2 \tpar f \nu^4 / 3 + \dots \, ,
\end{align}
and so,
\begin{align}
	E(M) - E(0) = \frac{2\pi^2 \tpar}{3} \nu^3 M+ \dots \, ,
\end{align}
Hence, the hopping yields an energy cost which is extensive in the number of improper bonds. 

Turning on $t_\perp$ allows for spin flips, provided they do not violate the Rydberg blockade. However, in the dilute particle limit, the average spacing of particles, $1/\nu$, becomes much larger than the blockade radius (of one lattice site), and thus we may approximate the blockade-conditioned spin-flip term $t_\perp$ as an unconditioned spin-flip in flavor space. Hence, the effective Hamiltonian for the flavor degrees of freedom is given by,
\begin{align}
	H_f = J(\nu)\sum_a \sigma^z_a \sigma^z_{a+1} - t_\perp \sum_a \sigma_a^x \, ,
\end{align}
with $J(\nu) = \pi^2 \tpar \nu^3 / 3$. This is the usual Ising Hamiltonian, which is critical at $t_\perp = J(\nu) = \pi^2 \tpar \nu^3 / 3$. This yields the small-filling limit of the transition line shown in Fig.~\ref{fig:igSPT}. Moreover, this analysis suggests the transition should be a $c= \frac{1}{2} + 1 = \frac{3}{2} $ conformal field theory, arising from the critical Ising flavor sector and the spectator gapless itinerant sector.

\begin{figure}
    \centering
    \includegraphics[width=\linewidth]{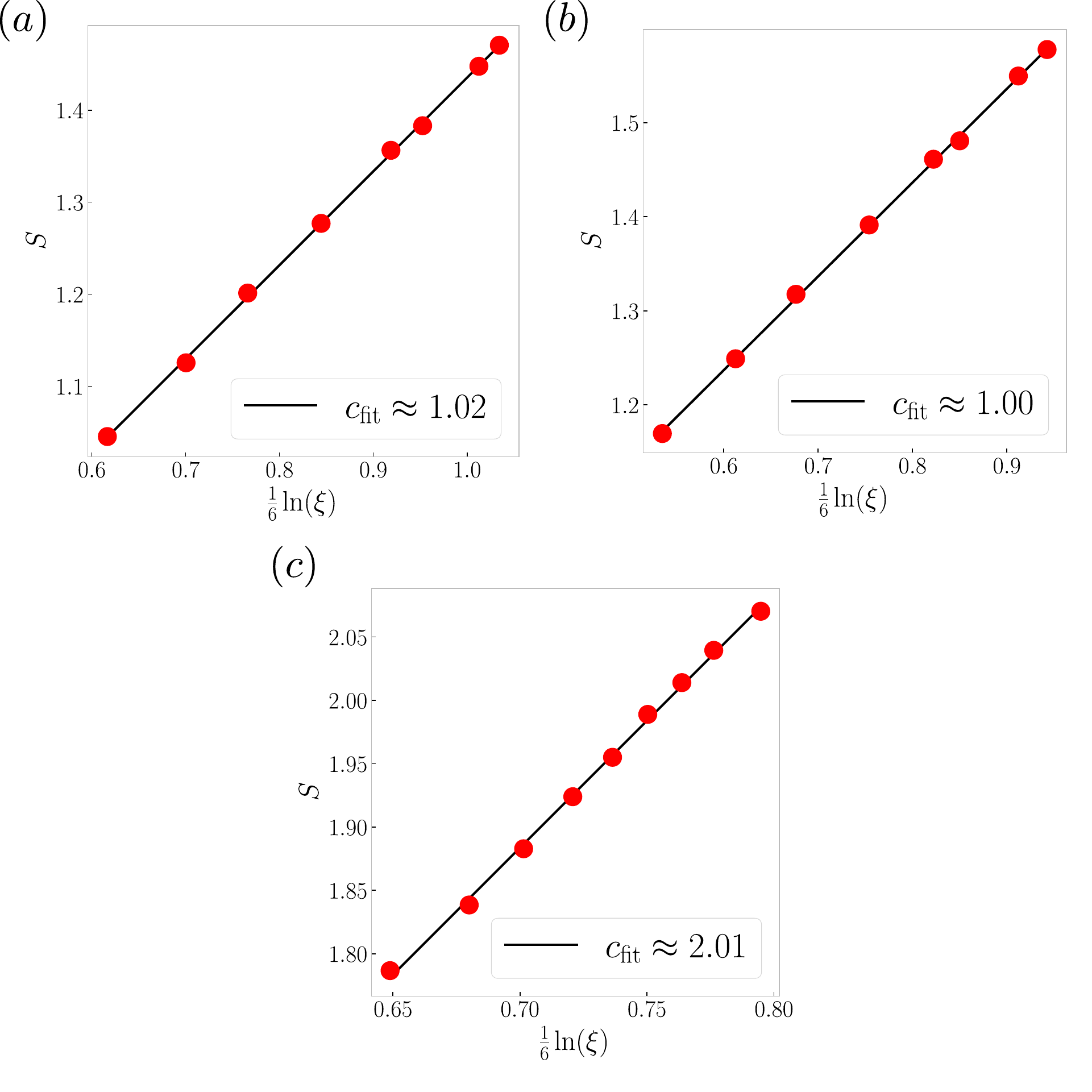}
    \caption{Finite-entanglement scaling to extract the central charge, $c_{\mathrm{fit}}$, for the two-leg model (a) in the trivial phase with $t_{\parallel,\perp}=1$ and $\nu=1/3$ and (b) the topological phase with $t_{\parallel,\perp}=1$ and $\nu=2/3$, as well as (c) the three-leg model with $\tper = 0.2\tpar$ and $\nu=6/7$.}
    \label{fig:c-extraction}
\end{figure}

\textit{Central charge---}We present in Fig.~\ref{fig:c-extraction} the results of the finite-entanglement scaling \cite{Pollmann2009} used to extract the central charge in the different gapless phases discussed in the main text. We see the extracted central charges are consistent with a single component Luttinger liquid ($c=1$) in the trivial and topological gapless phases of the two-leg model and a two component Luttinger liquid ($c=2$) in the three-leg ladder.

\textit{Supersymmetry---}Here, we establish the equivalence of Eq.~\eqref{eq:PXP_Eff} (up to the chemical potential), to the supersymmetric (SUSY) model studied in Ref.~\onlinecite{Bauer2013}. There, the authors study a two-leg fermion chain with a lattice SUSY, described by the Hamiltonian,
\begin{align}
    H_c = t\sum_{\langle i,j\rangle} P \left( c_i^\dagger c_j + h.c. \right) P + t\sum_i P_i, \label{eq:H_f}
\end{align}
where the $c_i$ are fermion annihilation operators while the $P$ and $P_i$ operators are defined as above. 
To emphasize the similarity, we here write Eq.~\eqref{eq:PXP_Eff} in terms of hardcore bosons, dropping the chemical potential:
\begin{align}
    H_a = -t\sum_{\langle i,j\rangle} P \left( a_i^\dagger a_j + h.c. \right) P + t\sum_i P_i \label{eq:H_b} \, .
\end{align}
We see that Eqs.~\eqref{eq:H_f} and \eqref{eq:H_b} only differ in that one is defined in terms of fermions and the other bosons. (The relative sign between the hopping and potential terms is readily eliminated via a gauge transformation using the bipartite structure of the two-leg ladder.) Remarkably, the blockade on the two-leg geometry results in these models being equivalent.

To establish this, let us use the pseudospin notation for the two-leg model and work with open boundary conditions. Then we define a zig-zag ordering of the sites: $(1,\downarrow) < (1,\uparrow) < (2,\downarrow) < (2,\uparrow) < \dots (L,\downarrow) < (L,\uparrow)$. We use this string to define Jordan-Wigner fermions,
\begin{align}
    c_{j} = \exp\left(i\pi \sum_{k<j} n_k  \right) a_{j} ,
\end{align}
where $\downarrow < \uparrow$ and $j$ indexes the sites as per the above ordering. Then, due to the hardcore constraint, the fermionic tunneling term for $i<j$ becomes,
\begin{align}
    c_i^\dagger c_j = a_i^\dagger \left[ \prod_{i<l<j} (1-2n_l) \right] a_j \, .
\end{align}
Now, for nearest-neighbor  sites $i$ and $j$, we in fact have that the Jordan-Wigner string is \emph{trivial},
\begin{align}
    P(c_i^\dagger c_j + h.c.)P = P(a_i^\dagger a_j + h.c.)P \, .
\end{align}
Indeed, there are two cases: $i$ and $j$ lie on the same rung, or on the same leg. If $i$ and $j$ are on the same rung, there are no intervening sites, and so the string trivially equals unity. If $i$ and $j$ are on the same leg, exactly one site in the ordering defined above separates them. However, this site must be on the same rung as one of $i$ and $j$, and thus must be unoccupied due to the blockade for an allowed hopping, such that the string equals unity. Hence, Eqs.~\eqref{eq:H_f} and \eqref{eq:H_b} are indeed equivalent, establishing the lattice SUSY of the latter.

\clearpage

\onecolumngrid

\begin{center}
{\large\bfseries Supplemental Material}
\end{center}
\setcounter{equation}{0}
\counterwithout{equation}{section}
\renewcommand{\theequation}{S\arabic{equation}}
\setcounter{section}{0}
\renewcommand{\thesection}{\Alph{section}}
\setcounter{subsection}{0}
\renewcommand{\thesubsection}{\thesection.\arabic{subsection}}
\setcounter{secnumdepth}{2}

\section{Energetic Decoupling of Itinerant and Internal Degrees of Freedom}

Let us consider the $N$-leg PXYP Hamiltonian, which we reproduce here: 
\begin{subequations}
\begin{align}
	\tag{\theparentequation} H &= H_\parallel + H_\perp \\
	H_\parallel &= - \tpar \sum_{i,\alpha} P \left( b_{i+1,\alpha}^\dagger b_{i,\alpha} + h.c. \right) P \\
	H_\perp &= - \tper \sum_{i,\alpha} P \left( b_{i,\alpha+1}^\dagger b_{i,\alpha} + h.c.  \right) P \, .
\end{align}
\end{subequations}
We will restrict ourselves to  $N=2,3$. In $H_\perp$, each rung bond is only counted once. In this section, we demonstrate quantitatively the sense in which the Rydberg constraint induces an energetic decoupling between the itinerant and internal degrees of freedom of the $N$-leg PXYP model.
As described in the main text, the physics of this model is most clearly expressed in squeezed space. Let us work in the particle occupation basis and label the positions of each particle as,
\begin{align}
	x_1 < x_2 < \dots < x_{N_p}
\end{align}
where $N_p$ is the number of particles. This ordering is well-defined as the blockade prevents particles from passing one another for $N=2,3$. We label the flavor of the $j^{th}$ particle with $q_j$. We then define the number of holes between particle $j$ and $j+1$ to be,
\begin{align}
	g_j = x_{j+1} - x_j - 1 \geq 0 \, .
\end{align}
The Rydberg constraint then becomes,
\begin{align}
	q_j = q_{j+1} \implies g_j > 0 \, .
\end{align}
That is, matching sequential flavors necessarily trap at least one hole. We further define,
\begin{align}
	d_j = \begin{cases}
		0 , \quad q_j \neq q_{j+1} \\
		1, \quad q_j = q_{j+1} \, .
	\end{cases}
\end{align}
We call a sequence of flavors $W = (q_1, q_2, \dots , q_{N_p})$ a \emph{word} and call $W$ a \emph{proper word} if $d_j \equiv 0$ for all $j$.

\subsection{Maximal Filling}

Let us first consider the case of maximal filling, such that there are no hole dynamics. Indeed, the dynamics of the system is determined entirely by $H_\perp$ in the subspace of proper words, with $N_p = L$. Interpreting the flavor $q_j$ as a height field, we define the gradient field, 
\begin{align}
	\eta_j \equiv q_{j+1} - q_j \,\, \text{mod} \, N .
\end{align}
In the proper word space, we have that $\eta_j \neq 0$ for all $j$. Now, the change of a flavor on a site $j$ induced by the hopping of $H_\perp$ results in,
\begin{align}
	q_j \mapsto q_{j} \pm 1 \implies (\eta_{j-1} , \eta_{j}) \mapsto (\eta_{j-1} \pm 1, \eta_{j} \mp 1) \, .
\end{align}
Hence, there is a $U(1)$ symmetry generated by the winding number,
\begin{align}
	Q_f \equiv \sum_{j} (\eta_j-1) \equiv \sum_{j} n_j^f \, , \label{eq:emergent_U1}
\end{align}
where we have defined $n_j^f = \eta_j - 1 \in \{ 0 , \dots , N-2\}$.
Note that, with periodic boundary conditions, we have that $Q_f = -L \,\, \text{mod}\, N$. The key physical point is that fluctuations in this winding charge are gapless, yielding a single $c=1$ mode for finite $\tper$, for $N = 3$; for $N=2$, the flavor sector has two word states and hence is pinned to an antiferromagnetic configuration, as is familiar from the Ising-Hubbard model. 

To be more explicit, focusing on the $N=3$ case, we can define the bosonic operators, $B_j$ and $B_j^\dagger$, with action on the flavor gradient fields,
\begin{align}
	B_j^\dagger \ket{n} = 
	\begin{cases}
		\ket{n+1} , \quad & n = 0 \\
		0 , \quad & n = 1
	\end{cases}, \quad
		B_j \ket{n} = 
	\begin{cases}
		\ket{n-1} , \quad & n = 1 \\
		0 , \quad & n = 0
	\end{cases}
\end{align}
Using these operators, we can write in the maximal filling subspace,
\begin{align}
	H_\perp = - \tper \sum_j (B_j^\dagger B_{j+1} + h.c. ) \, .
\end{align}
This is the $U(1)$-symmetric XY Hamiltonian. 
In particular, since the system is symmetric under inversion of flavors, i.e. $q_j \mapsto -q_j$, we have that the filling is fixed to be, for $N=3$, $Q_f / L = 1/2$. For $N=3$, the flavor sector is thus at half-filling and realizes a Luttinger liquid at low energies. 

\subsection{Small Hole Doping}

We now consider the limit of close to maximal filling, 
such that the density of holes is small, but finite in the thermodynamic limit.
We wish to argue that, in this limit, the hole and flavor sectors are effectively decoupled, such that we get a two component Luttinger liquid, with the two components arising from the itinerant holes and the gapless flavor sector.
Specifically, the claim is that the \emph{proper word} sector, as defined above, remains the low energy manifold, serving as a background on top of which the holes hop. The physical picture behind this is that in any word sector, an improper flavor pair (i.e. two particles that are nearest-neighbor in squeezed space with the same flavor) binds one hole, such that the total number of mobile holes is decreased. This has the effect of the original particles hopping on an effectively smaller lattice, which costs kinetic energy.

Explicitly, let us define,
\begin{align}
	M = \sum_j d_j,
\end{align}
to count the total number of improper bonds (i.e. a neighboring pair of particles in squeezed space with the same flavor)
where we recall that $d_j = 1$ for an improper bond and $d_j=0$ otherwise. Then, we have that $g_j \geq d_j$. Moreover, defining
\begin{align}
	\tilde{g}_j = g_j - d_j,
\end{align}
we have that,
\begin{align}
	\sum_j g_j = M + \sum_j \tilde{g}_j \, .
\end{align}
This expresses the fact that $M$ of the holes are bound to improper bonds. 

Let us consider the limit with $\tper=0$ such that the word is conserved. If $M=0$, it is plain to see that the $x_j$ describe hardcore bosons with a nearest neighbor hopping on an $L$ site chain (let us assume open boundary conditions). The claim is that for finite $M$, the system is equivalent to hardcore bosons with nearest-neighbor hopping on a reduced chain of length $L-M$. To see this, let us define,
\begin{align}
	D_j = \sum_{k<j} d_k
\end{align}
as the number of improper bonds to the left of the particle $j$ and set
\begin{align}
	y_j = x_j - D_j.
\end{align}
This shift accounts for the holes bound to improper bonds. 
Now,
\begin{align}
	y_{j+1} - y_j = 1 + \tilde{g}_j \, .
\end{align}
Hence, $y_{j+1} > y_j$, and thus the $y_j$ correspond to the coordinates of hardcore bosons. Moreover, these hardcore bosons also obey nearest-neighbor hopping. Indeed, a shift of $x_j$ to $x_j \pm 1$ implies a shift of $y_j \to y_j \pm 1$ since, in a given word sector, all the $d_j$ and hence the $D_j$ are fixed. Recall that $N_p$ is the total number of particles. The maximum value $y_{N_p}$ can take is $L-M$. So, the $y_j$ describe hardcore bosons hopping on a chain of reduced length $L-M$.

Now, via the Jordan-Wigner transformation, the dispersion of hardcore bosons is given by $\epsilon(k) = -2 \tpar \cos(k)$ (where we set the lattice spacing to be $a=1$). If the number of these bosons (which is also the number of physical bosons) is $N_p$, the Fermi wavevector is given by $k_F = \pi N_p/ L_{\mathrm{eff}} = \pi N_p /(L-M)$, where $L_{\mathrm{eff}} = L-M$ is the effective length in the word sector with $M$ improper bonds. Thus, in the thermodynamic limit, the total energy of the ground state in the $M$ improper bond sector is given by,
\begin{align}
	E(M) = L_{\mathrm{eff}} \int_{-k_F}^{k_F} \frac{dk}{2\pi} \epsilon(k) = -\frac{2 \tpar (L-M)}{\pi} \sin \left( \frac{\pi N_p}{L-M} \right) \, .
\end{align}
Setting $\nu = N_p/L$ to be the particle filling and $\delta = 1 - \nu$ the hole filling, we find that the approximate energy cost to excite an improper bond above the proper word subspace is, in the small hole doping limit,
\begin{align}
	\Delta E \approx \left. \frac{dE(M)}{dM}\right|_{M=0} &= 2\tpar - 2\tpar \pi^2\delta^2 /2 + O(\delta^3) \, .
\end{align}
Hence, in the limit of small doping, adding a single improper bond costs an $O(1)$ energy of approximately $2\tpar$; adding a sufficiently dilute number of improper bonds should then cost an extensive (in the number of added improper bonds) amount of energy. This suggests that at low hole dopings (i.e. close to maximal filling), we should expect the dynamics to effectively be constrained to the proper word space, even when we turn on a small flavor coupling, $\tper$. So, the dynamics of the holes and the flavor degrees of freedom are effectively decoupled. 

For the two-leg ladder this implies we should get a single-component Luttinger liquid at small hole doping, coming from the itinerant degrees of freedom, while the frozen squeezed space degrees of freedom yield long-range string order. For the three-leg ladder, this implies that we should expect a two-component Luttinger liquid, i.e. a conformal field theory with central charge $c=2$, accounting for both the hole and flavor dynamics, in the case of the three-leg ladder. 

\section{Bosonized Stability Analysis}

Let us focus on the $N=3$ case and study the stability of the $c=2$ phase. Starting from maximal filling, we can assume the itinerant and internal degrees of freedom remain decoupled at low hole doping. The itinerant degrees of freedom thus give another Luttinger liquid. The goal is to check the stability of the internal Luttinger liquid to doping.

We use the standard bosonization map \cite{Giamarchi2004},
\begin{align}
	n_f(x) &\sim \bar n_f - \frac{1}{\pi} \partial_x \phi_f + \sum_{r \geq 1} A_r \cos\left[ 2r(\pi \bar n_f x - \phi_f) \right] \\
	B^\dagger(x) &\sim e^{-i\theta_f} \sum_r C_r e^{2ir (\pi \bar n_f x - \phi_f)} \, .
\end{align}
Here, $\bar n_f = 1/2$ as above. The internal degrees of freedom are described at low energy by the Hamiltonian,
\begin{align}
	H_f = \frac{v_f}{2\pi} \int dx \left[ K_f(\partial_x \theta_f)^2 + K_f^{-1} (\partial_x \phi_f)^2 \right] \, .
\end{align}
The velocity $v_f$ and the Luttinger parameter $K_f$ are set by microscopic parameters. We use a normalization such that the vertex operator $V_{m,r} = \exp\left(i(m\theta_f + 2r\phi_f) \right)$ has scaling dimension $\Delta_{m,r} = m^2 (4K_f)^{-1} + r^2 K_f$. 

We wish to identify instabilities of this Luttinger liquid phase. As usual, one potential instability arises from Umklapp scattering. Since $\bar n_f = 1/2$ for $N=3$, the lowest harmonic Umklapp scattering process is given by the operator $g_U \cos(4\phi_f)$, which has scaling dimension $\Delta_U = \Delta_{0,2}=4K_f$, and is thus irrelevant for $K_f > 1/2$. Note that this instability is present even at maximal filling. 

When we dope the system with holes, virtual processes involving improper bonds can eventually break the emergent $U(1)$ symmetry generated by $Q_f$ [Eq.~\eqref{eq:emergent_U1}]. Indeed, for $N=3$, consider a configuration such that, for three consecutive bonds in squeezed space, at least one can take the value $\eta_j = 0$ (yielding an improper bond). Then a virtual process like the following can occur:
\begin{align}
	(a,1,c) \to (a+1, 0 , c) \to (a+1, N-1, c+1) ,
\end{align}
with $a,c \in \{ 0 , \dots , N-2\}$. If $a$ and $c$ are such that the initial configuration is a proper word, then the final one is as well but $Q_f$ has increased by $N=3$. Hence, for finite hole doping, virtual processes can change $Q_f$ but the proper word constraint implies $\Delta Q_f \in N \mathbb{Z}$. In the effective field theory description, this implies we can include the perturbation,
\begin{align}
	H_\lambda = \int dx \, \lambda \cos(N \theta_f) \, .
\end{align}
Since such virtual processes require the presence of holes, we have $\lambda \propto \delta$. Although $\cos(N \theta_f)$ should therefore also couple to fluctuations in the hole density, $\delta(x)$, such terms are necessarily less relevant and can be ignored. We have that $\Delta_\lambda = N^2 (4K_f)^{-1}$, and hence $H_\lambda$ is irrelevant for $K_f < N^2 / 8$. 

For the case of interest, $N=3$, we see that both instabilities are irrelevant if
\begin{align}
	\frac{1}{2} < K_f < \frac{9}{8} \, .
\end{align}
Hence, there is a finite window of stability of the two-component Luttinger liquid at finite hole density. The numerical results presented in the main text thus appear to lie within this window.
This two-component Luttinger liquid picture also explains the difference in decay of the conventional correlator and string order in Fig.~\ref{fig:igSPT}(c) of the main text. In this bosonized picture, the string order only depends on the Luttinger parameter for the gapless flavor degrees of freedom, whereas the conventional correlator depends on the Luttinger parameters for both the flavor and itinerant degrees of freedom.

\end{document}